\documentclass[12pt,a4paper]{article}
\usepackage[utf8]{inputenc}
\usepackage[T1]{fontenc}
\usepackage{amsmath}
\usepackage{amsfonts}
\usepackage{amssymb}
\usepackage{cite} 
\usepackage{color}
\usepackage[title]{appendix}
\usepackage{geometry}
\DeclareMathOperator{\sech}{sech}
\newcommand{\appref}[1]{Appendix~\ref{#1}}
\usepackage{hyperref}
\usepackage{cleveref}
\hypersetup{colorlinks=true}

\title{\textbf{BCS Gap Equation on Riemannian Manifolds: A Heat Kernel Analysis}}
\author{
  Levent Akant\thanks{\texttt{levent.akant@bogazici.edu.tr}}
  \and
  Emine Ertu\u{g}rul\thanks{\texttt{emine.ertugrul@bogazici.edu.tr}}
  \and
  O. Teoman Turgut\thanks{\texttt{turgutte@bogazici.edu.tr}}
}
\date{
  \normalsize \textit{Department of Physics, Bo\u{g}azi\c{c}i University, Bebek, Istanbul, T\"{u}rkiye}\\[2ex]
  \today
}

\begin{document}

\maketitle

\begin{abstract}
In this paper, we investigate the Bardeen-Cooper-Schrieffer (BCS) theory of superconductivity formulated on a three dimensional Riemannian manifold. We derive the gap equation in the path integral picture by employing a Hubbard-Stratonovich transformation followed by a saddle point approximation. Under the assumption that the length scale set by the metric is much larger than the healing length, we solve the gap equation to find a slowly varying gap function that exhibits the effects of curvature. To rigorously address the ultraviolet divergences inherent in the gap equation, we utilize heat kernel expansion techniques, which provide a systematic and mathematically transparent renormalization scheme. We explicitly evaluate the renormalized gap equation at both zero and finite temperatures, computing the relevant integrals asymptotically in the weak-coupling limit ($\Delta \ll \mu$). Furthermore, we establish a universal, regularization-independent relation connecting the finite-temperature gap to the zero-temperature gap and the temperature itself. Based on this fundamental relation, we analytically prove the monotonic decrease of the energy gap with increasing temperature ($\partial \Delta / \partial T < 0$), successfully recovering the standard universal BCS behavior near the critical temperature within a curved space framework.
\end{abstract}

\section{Introduction}

It is well known that interactions between fermions lead to a change in the Fermi surface; 
for most weak repulsive interactions, this can be understood in terms of quasi-particle excitations.
However, once the interactions become attractive, the Fermi surface develops an instability 
and a pairing mechanism leads to a decrease in the ground state energy, which is not 
perturbatively connected to the free Fermi surface. More precisely, the new ground state cannot 
be represented in the Fock space of the free theory as a normalizable state. 
This phenomenon is the basis of superconductivity, and the phonon exchange mechanism near the 
Fermi surface leads to a simple attractive interaction, which is the main idea of BCS theory \cite{bcs1, bcs2, cooper}. 
It is believed that the mechanism behind this is universal and mostly independent of the specific form of the interaction. This point can be understood from a renormalization group perspective; a good reference for this is the review article by Shankar \cite{shankar-rev} and his book on this matter \cite{shankar-book}.

In curved space, the renormalization group point of view is not fully developed \cite{birrell, parker, dewitt, inagaki}; 
nevertheless, it is a good approximation to assume that typical wavelengths near the Fermi surface are much smaller than the length scale defined by the curvature. That does not 
mean that the curvature is unimportant since the resulting partition function depends on 
the spectrum. Indeed, recent literature has seen renewed interest in exploring fermion condensation and BCS-like mechanisms in curved backgrounds, such as inflationary spacetimes \cite{tong2024}. Concurrently, there have been significant advances in fermionic heat kernel techniques, including calculations for nonzero spin in curved spacetimes \cite{carneiro2025}, resummation methods for spinor fields \cite{franchino2026}, and evaluations of geometric effective actions \cite{gattus2024, alonso2019}. In this work, we investigate the gap equation of the BCS theory using a detailed heat kernel analysis.

We introduce a simple model Hamiltonian of non-relativistic fermions on a Riemannian manifold with metric $g$,
\begin{equation}
H = \int  d_g x\; \bar{\psi}_{\sigma}
    \left[-\frac{\hbar^2 }{2m}\left(\nabla^2-\xi R\right)\right]\psi_{\sigma} 
    - u_0 \int  d_g x\; \overline{\psi}_{\uparrow}\overline{\psi}_{\downarrow} \psi_{\downarrow}\psi_{\uparrow},
\label{eq:Hamiltonian}
\end{equation}
where $d_g x=d^3 x \sqrt{g(x)}$ is the Riemann volume element, $\nabla^2$ is the Laplacian corresponding to $g$, $R$ is the scalar curvature, and repeated spin indices $\sigma$ are summed. This model, which is obtained by coupling the flat space theory to the metric in a standard way, may seem naive for describing superconductivity in a curved space; nevertheless, we believe it is a good toy model to develop the required mathematical techniques and to carefully understand the curvature effects on such systems. 

It is well-known that the partition function of an interacting theory can be elegantly expressed 
in a path integral formalism as \cite{stoof}
\begin{equation}
    Z = \int_{\text{apbc}} \mathcal{D}\bar{\psi} \mathcal{D}\psi \, e^{-\frac{1}{\hbar}S[\bar{\psi},\psi]}
\end{equation}
where for fermions, we use the anti-periodic boundary conditions (apbc) at imaginary periodic time. We will assume that the system is weakly interacting, staying away from the crossover regime. To evaluate this partition function, we will employ a Hubbard-Stratonovich transform \cite{hubbard, stratonovich} to decouple fermionic interactions. The gap equation will be derived as the saddle point equation of the resulting effective action.  

In curved space, we will assume that the scale $\ell_g$ over which the geometry changes appreciably is much larger than the healing length $\ell_h$ of the order parameter. Under this assumption, we will take a locally uniform condensate, thereby adopting the strategy underlying the local density approximation \cite{KohnSham1965} used in the analysis of superconductivity in the presence of a slowly varying trapping potential \cite{Perali2003, Perali2004a, Perali2004b, GiorginiPitaevskiiStringari2008, Bulgac2007}. This will enable us to solve the BCS gap equation for a locally constant (i.e. slowly varying, in the spirit of the local density approximation) fermion condensate $\Delta(x)$, which will nonetheless retain some curvature effects. 

Going beyond the local density approximation requires a derivative (gradient) expansion of the underlying operators about a base point (\textit{e.g.} in Riemann normal coordinates). In the present work, we  will only attempt the zeroth-order calculation. Even under this simplifying assumption, we find that many subtleties arise and the calculations must be performed with great care.

The organization of this paper is as follows. In Section 2, we introduce the model Hamiltonian for non-relativistic fermions on a background  Riemannian manifold and derive the gap equation using the path integral formalism employing a Hubbard-Stratonovich transformation under a uniform saddle-point approximation in a slowly varying geometry. In Section 3, we present the heat kernel expansion method for the relevant differential operators, systematically organizing the expansion terms using parabolic cylinder functions. In Section 4, we utilize this heat kernel framework to regularize and then renormalize the gap equation at zero as well as at finite temperatures. In particular, we determine the curvature effects on the slowly varying gap function $\Delta(x)$, derive the critical temperature, relate the zero and finite temperature gap solutions, analyze the behavior of the gap near $T_c$, and establish a universal, cutoff-independent relation that governs the monotonic decrease of the gap with increasing temperature. Section 5 contains our conclusions and outlook. Finally, the appendices (A through D) provide  detailed derivations of the mathematical identities, the asymptotic evaluation of the integrals involved, and the explicit calculation of the thermodynamic derivatives required for our analysis.

\section{BCS Gap Equation in a Curved Manifold}

In this section, we first derive the exact gap equation on a Riemannian manifold using the path integral formalism, and then simplify it by applying a geometric version of the local density approximation.

\subsection{Derivation of the Gap Equation}

The partition function for the BCS theory on a Riemannian manifold with metric tensor $g_{ij}(x)$ is given  as
\begin{equation}
 Z=\int\mathcal{D}\psi \mathcal{D}\overline{\psi}\,e^{-\frac{1}{\hbar}S[\overline{\psi},\psi]},
\end{equation}
with
\begin{equation}
 S[\overline{\psi},\psi]=\int_{\tau,x}\,\sqrt{g(x)}\,\left\{\overline{\psi}_{\sigma}\left[\hbar\partial_{\tau}-\frac{\hbar^{2}\nabla^{2}}{2m}+U(x)-\mu\right]\psi_{\sigma}-u_{0}\overline{\psi}_{\uparrow}\overline{\psi}_{\downarrow}
 \psi_{\downarrow}\psi_{\uparrow}\right\},
\end{equation}
where $\nabla^2$ is the Laplace-Beltrami operator, 
\begin{equation}
  \nabla^{2}=\frac{1}{\sqrt{g(x)}}\partial_{i}\sqrt{g(x)}g^{ij}(x)\partial_{j},
\end{equation}
$U(x)$ is a potential term, which includes the coupling to scalar curvature, and the fermion measure is given as
\begin{equation}
 \mathcal{D}\psi \mathcal{D}\overline{\psi}=\prod_{\sigma,\tau,x}d\left[g^{1/4}(x)\psi_{\sigma}(\tau,x)\right]d\left[g^{1/4}(x)\overline{\psi}_{\sigma}(\tau,x)\right].
\end{equation}
Here and in what follows, we use the shorthand notation
\begin{equation}
  \int_{\tau,x}=\int_{0}^{\beta\hbar}d\tau\int d^{3}x.
\end{equation}
For later use, we integrate by parts the term in $S$ which is quadratic in spin down fields and write
\begin{equation}
S[\overline{\psi},\psi]=\int_{\tau,x}\,\sqrt{g(x)}\,\left\{\overline{\psi}_{\uparrow}h_{1}\psi_{\uparrow}
 +\psi_{\downarrow}h_{2}\overline{\psi}_{\downarrow}
 -u_{0}\overline{\psi}_{\uparrow}\overline{\psi}_{\downarrow}
 \psi_{\downarrow}\psi_{\uparrow}\right\}.
\end{equation}
where
\begin{align}
  h_{1} &= \hbar\partial_{\tau}-\frac{\hbar^{2}\nabla^{2}}{2m}+U(x)-\mu \\
  h_{2} &= \hbar\partial_{\tau}+\frac{\hbar^{2}\nabla^{2}}{2m}-U(x)+\mu=-h_{1}^{\dagger}
\end{align}
Although our method works for any slowly varying potential $U(x)$, from here on we will take it to be the curvature coupling
\begin{equation}
    U(x)=\frac{\hbar^2}{2m}\xi R(x).
\end{equation}
The field redefinition
\begin{equation}
  \phi_{\sigma}(\tau,x)=g^{1/4}(x)\psi_{\sigma}(\tau,x),\;\;\;\;\; \overline{\phi}_{\sigma}(\tau,x)=g^{1/4}(x)\overline{\psi}_{\sigma}(\tau,x)
\end{equation}
gives
\begin{equation}\label{Parti}
 Z=\int\mathcal{D}\phi \mathcal{D}\overline{\phi}\,\exp\left[\frac{1}{\hbar}\int_{\tau,x}\,\overline{\phi}_{\uparrow}L_{1}\phi_{\uparrow}+
 \phi_{\downarrow}L_{2}\overline{\phi}_{\downarrow}\right]
 \exp\left\{\frac{u_{0}}{\hbar}\int_{\tau,x}\,[g^{-1/4}\overline{\phi}_{\uparrow}\overline{\phi}_{\downarrow}][g^{-1/4}\phi_{\downarrow}\phi_{\uparrow}]\right\},
\end{equation}
where
\begin{equation}
  L_{1}=-g^{1/4}(x)h_{1}g^{-1/4}(x),\;\;\;\;\;L_{2}=-g^{1/4}(x)h_{2}g^{-1/4}(x),
\end{equation}
and
\begin{equation}
 \mathcal{D}\phi \mathcal{D}\overline{\phi}= \prod_{\sigma,\tau,x}\left[d\phi_{\sigma}(\tau,x)d\overline{\phi}_{\sigma}(\tau,x)\right].
\end{equation}
Consider the Hubbard-Stratonovich transform
\begin{equation}
 \exp\left\{\frac{u_{0}}{\hbar}\int_{\tau,x}\,[g^{-1/4}\overline{\phi}_{\uparrow}\overline{\phi}_{\downarrow}][g^{-1/4}\phi_{\downarrow}\phi_{\uparrow}]\right\}=\mathcal{N}\int\mathcal{D}F \mathcal{D}\overline{F} \exp\left\{-\frac{1}{\hbar}\int_{\tau,x}\left[\frac{|F|^{2}}{u_{0}}-\overline{F}g^{-1/4}\phi_{\downarrow}\phi_{\uparrow}-Fg^{-1/4}\overline{\phi}_{\uparrow}\overline{\phi}_{\downarrow} \right] \right\}
\end{equation}
where $\mathcal{N}$ is a normalization constant and $F$ is a complex scalar field with
\begin{equation}
  \mathcal{D}F \mathcal{D}\overline{F} =\prod_{\tau,x}\,dF(\tau,x)d\overline{F}(\tau,x).
\end{equation}
Now, performing the field redefinition
\begin{equation}
  \Delta(\tau,x)=g^{-1/4}(x)F(\tau,x),\;\;\;\; \bar{\Delta}(\tau,x)=g^{-1/4}(x)\overline{F}(\tau,x)
\end{equation}
we get
\begin{equation}\label{HS}
 \exp\left\{\frac{u_{0}}{\hbar}\int_{\tau,x}\,[g^{-1/4}\overline{\phi}_{\uparrow}\overline{\phi}_{\downarrow}][g^{-1/4}\phi_{\downarrow}\phi_{\uparrow}]\right\}=\mathcal{N}\int\mathcal{D}\Delta \mathcal{D}\bar{\Delta} \exp\left\{-\frac{1}{\hbar}\int_{\tau,x}\left[\frac{\sqrt{g}|\Delta|^{2}}{u_{0}}-\bar{\Delta}\phi_{\downarrow}\phi_{\uparrow}-\Delta\overline{\phi}_{\uparrow}\overline{\phi}_{\downarrow}\right] \right\}
\end{equation}
with
\begin{equation}
  \mathcal{D}\Delta \mathcal{D}\bar{\Delta} =\prod_{\tau,x}\,[g^{1/4}(x)d\Delta(\tau,x)][g^{1/4}(x)d\bar{\Delta}(\tau,x)].
\end{equation}
Using \eqref{HS} in \eqref{Parti} and introducing the Nambu fields \cite{nambu}
\begin{equation}
  \Phi(\tau,x)=\left(
\begin{array}{c}
 \phi_{\uparrow}(\tau,x) \\
\overline{\phi}_{\downarrow}(\tau,x) \\
\end{array}
 \right),\;\;\;\;\overline{\Phi}(\tau,x)=\left(\begin{array}{cc}
 \overline{\phi}_{\uparrow}(\tau,x) & \phi_{\downarrow}(\tau,x)  \\
 \end{array}
 \right)
\end{equation}
we arrive at the following expression for the partition function
\begin{equation}
  Z=\mathcal{N}\int \mathcal{D}\Delta \mathcal{D}\bar{\Delta}\mathcal{D}\phi \mathcal{D}\overline{\phi}\,\exp\left[\frac{1}{\hbar}\int_{\tau,x} \overline{\Phi}M \Phi
  \right]\exp\left[-\frac{1}{\hbar u_{0}}\int_{\tau,x} \sqrt{g}\,|\Delta|^{2}\right].
\end{equation}
where
\begin{equation}
  M=\left(
      \begin{array}{cc}
        L_{1} & \Delta \\
        \bar{\Delta} & L_{2} \\
      \end{array}
    \right).
\end{equation}
Integration over the fermionic fields then yields
\begin{equation}
  Z=\mathcal{N}\int \mathcal{D}\Delta \mathcal{D}\bar{\Delta}\,\exp\left[\log\det \frac{M}{\hbar}-\frac{1}{\hbar u_{0}}\int_{\tau,x} \sqrt{g}\,|\Delta|^{2}\right].
\end{equation}
Note that the inverse of $M$ is given as,
\begin{equation}
  M^{-1}=\left(
      \begin{array}{ccc}
        -\bar{\Delta}^{-1}L_{2}(\Delta-L_{1}\bar{\Delta}^{-1}L_{2})^{-1} && (\bar{\Delta}-L_{2}\Delta^{-1}L_{1})^{-1} \\
         & & \\
         (\Delta-L_{1}\bar{\Delta}^{-1}L_{2})^{-1}&& -\Delta^{-1}L_{1}(\bar{\Delta}-L_{2}\Delta^{-1}L_{1})^{-1} \\
      \end{array}
    \right).
\end{equation}
In the path integral picture the mean field theory corresponds to the evaluation of the partition function in the saddle point approximation \cite{stoof}. Thus, varying the action with respect to $\Delta$ and $\bar{\Delta}$ and using the well known identity
\begin{equation}
  \delta\log\det M=\textrm{Tr}M^{-1}\delta M
\end{equation}
with the infinitesimal variation
\begin{equation}
  \delta M=\left(
      \begin{array}{cc}
        0 & \delta\Delta \\
        \delta\bar{\Delta} & 0 \\
      \end{array}
    \right)
\end{equation}
we get
\begin{align}\label{var1}
  \delta\left[\log\det \frac{M}{\hbar}-\frac{1}{\hbar u_{0}}\int_{\tau,x} \sqrt{g}\,|\Delta|^{2}\right] &= \left\{\textrm{Tr}\left[(\bar{\Delta}-L_{2}\Delta^{-1}L_{1})^{-1}\delta\bar{\Delta}\right]-\frac{1}{\hbar u_{0}}\int_{\tau,x}\sqrt{g(x)}\,\Delta(\tau,x)\delta\bar{\Delta}(\tau,x) \right.\nonumber\\
  &\quad+ \left.\textrm{Tr}\left[ (\Delta-L_{1}\bar{\Delta}^{-1}L_{2})^{-1}\delta\Delta\right]-\frac{1}{\hbar u_{0}}\int_{\tau,x}\sqrt{g(x)}\,\bar{\Delta}(\tau,x)\delta\Delta(\tau,x)\right\}.
\end{align}

For a differential operator $L$ defined on a Riemannian manifold the corresponding Green's function is given by the solution of \cite{avramidi-book}
\begin{equation}
  LG(x,x')=\delta(x,x'),
\end{equation}
where locally
\begin{equation}
  \delta(x,x')=\frac{\delta(x-x')}{\sqrt{g(x)}}=\frac{\delta(x-x')}{g^{1/4}(x)g^{1/4}(x')}.
\end{equation}
Then we have
\begin{equation}
  [g^{1/4}(x)Lg^{-1/4}(x)][g^{1/4}(x)G(x,x')g^{1/4}(x')]=\delta(x-x').
\end{equation}
In our case we have
\begin{equation}
\bar{\Delta}-L_{2}\Delta^{-1}L_{1}=g^{1/4}(x)\left[\bar{\Delta}-h_{2}\Delta^{-1}h_{1}
\right]g^{-1/4}(x)
\end{equation}
Let
\begin{equation}\label{Green12}
  \left[\bar{\Delta}-h_{2}\Delta^{-1}h_{1}
\right]G_{12}(\tau,x,\tau',x')=\delta(\tau-\tau')\delta(x,x').
\end{equation}
Then the integral kernel of $(\bar{\Delta}-L_{2}\Delta^{-1}L_{1})^{-1}$ is given by
\begin{equation}
  g^{1/4}(x)G_{12}(\tau,x,\tau',x')g^{1/4}(x').
\end{equation}
Consequently,
\begin{align}\label{trace1}
 \textrm{Tr}\left[(\bar{\Delta}-L_{2}\Delta^{-1}L_{1})^{-1}\delta\bar{\Delta}\right]&=\int_{\tau,x}\int_{\tau',x'}
 g^{1/4}(x)G_{12}(\tau,x,\tau',x')g^{1/4}(x')\delta\bar{\Delta}(\tau',x')\delta(\tau-\tau')\delta(x-x')\nonumber\\
 &=\int_{\tau,x} \sqrt{g(x)}\,G_{12}(\tau,x,\tau,x)\delta\bar{\Delta}(\tau,x).
\end{align}
Similarly,
\begin{equation}
  \Delta-L_{1}\bar{\Delta}^{-1}L_{2}=g^{1/4}(x)\left[\Delta-h_{1}\bar{\Delta}^{-1}h_{2}\right]g^{-1/4}(x),
\end{equation}
and
\begin{equation}\label{trace2}
 \textrm{Tr}\left[(\Delta-L_{1}\bar{\Delta}^{-1}L_{2})^{-1}\delta \Delta\right]=\int_{\tau,x} \sqrt{g(x)}\,G_{21}(\tau,x,\tau,x)\delta\Delta(\tau,x),
\end{equation}
where
\begin{equation}\label{Green2}
  \left[\Delta-h_{1}\bar{\Delta}^{-1}h_{2}\right]G_{21}(\tau,x,\tau',x')=\delta(\tau-\tau')\delta(x,x').
\end{equation}
Using \eqref{trace1} and \eqref{trace2} in \eqref{var1} we get
\begin{align}\label{var2}
  \delta\left[\log\det \frac{M}{\hbar}-\frac{1}{\hbar u_{0}}\int_{\tau,x} \sqrt{g}\,|\Delta|^{2}\right] &= \int_{\tau,x}\sqrt{g(x)}\,\left[G_{12}(\tau,x,\tau,x)-\frac{1}{\hbar u_{0}}\Delta(\tau,x)\right]\delta\bar{\Delta}(\tau,x) \nonumber\\
&\quad+\int_{\tau,x}\sqrt{g(x)}\,\left[G_{21}(\tau,x,\tau,x)-\frac{1}{\hbar u_{0}}\bar{\Delta}(\tau,x)\right]\delta\Delta(\tau,x).
\end{align}
Thus, we have
\begin{equation}\label{gap}
  \frac{\bar{\Delta}(\tau,x)}{\hbar u_{0}}= G_{21}(\tau,x,\tau,x),\qquad \frac{\Delta(\tau,x)}{\hbar u_{0}}= G_{12}(\tau,x,\tau,x).
\end{equation}

In this paper, we will consider only static, real valued gap functions. If we assume that the gap function is independent of $\tau$, our system becomes homogeneous in time. Consequently, the Green's functions become functions of $\tau-\tau'$ and we can expand
\begin{equation}
   G_{21}(\tau,x,\tau',x')=  \frac{1}{\hbar\beta} \sum_{j=-\infty}^{\infty} e^{-i\omega_j (\tau-\tau')} \, G_{21}(x,x';i\omega_j) 
\end{equation}
with a similar expansion for $ G_{12}(\tau,x,\tau',x')$. Here $\omega_j$'s are the Matsubara frequencies
\begin{equation}
    \omega_j = \frac{(2j+1)\pi}{\hbar\beta}, \qquad j \in \mathbb{Z}.
\end{equation}
Then the gap equations are given as
\begin{eqnarray}\label{gapfreq}
  \frac{1}{ u_{0}}&=& \frac{1}{\beta} \sum_{j=-\infty}^{\infty}  \bar{\Delta}^{-1}(x)\, G_{21}(x,x;i\omega_j),\\ \frac{1}{ u_{0}}&=&\frac{1}{\beta}  \sum_{j=-\infty}^{\infty}  \Delta^{-1}(x
  )\, G_{12}(x,x;i\omega_j).
\end{eqnarray}
In the frequency domain (\ref{Green12}) and (\ref{Green2}) can be expressed as
\begin{equation}\label{Green12fr}
  \left[\bar{\Delta}(x)\Delta(x)-h_{2}(i\omega_j)\Delta^{-1}(x)h_{1}(i\omega_j)\Delta(x)
\right]\left[\Delta^{-1}(x)\,G_{12}(x,x';i\omega_j)\right]=\delta(x,x').
\end{equation}
\begin{equation}\label{Green21fr}
  \left[\Delta(x)\bar\Delta(x)-h_{1}(i\omega_j)\bar{\Delta}^{-1}(x)h_{2}(i\omega_j)\bar{\Delta}(x)\right]\left[\bar{\Delta}^{-1}(x)G_{21}(x,x';i\omega_j)\right]=\delta(x,x').
\end{equation}
where
\begin{align}
  h_{1}( i \omega_{j}) &= i\hbar \omega_{j}-\frac{\hbar^{2}\nabla^{2}}{2m}+U(x)-\mu \\
  h_{2} (i \omega_{j}) &=  i\hbar \omega_{j}+\frac{\hbar^{2}\nabla^{2}}{2m}-U(x)+\mu=-h_{1}^{\dagger}( i \omega_{j}).
\end{align}
Note that $\bar{h}_{1}( i \omega_{j})=-h_{2}( i \omega_{j})$ and this, in turn, implies
\begin{equation}
   \overline{\left[\bar{\Delta}-h_{2}(i\omega_j)\Delta^{-1}h_{1}(i\omega_j)
\right]}= \left[\Delta-h_{1}(i\omega_j)\bar{\Delta}^{-1}h_{2}(i\omega_j)\right].
\end{equation}
Hence,
\begin{equation}
  \bar{G}_{12}(x,x',i\omega_j)=G_{21}(x,x',i\omega_j). 
\end{equation}
Also note that,
\begin{equation}\label{operator}
 |\Delta|^{2}-h_{1}(i\omega_j)\bar{\Delta}^{-1}h_{2}(i\omega_j)\bar{\Delta}=
  |\Delta|^{2}-h_{1}(i\omega_j)h_{2}(i\omega_j)-h_{1}(i\omega_j) \bar{\Delta}^{-1}\left[\frac{\hbar^{2}}{2m}\nabla^{2}\bar{\Delta}+
  \frac{\hbar^{2}}{m}(\nabla^{\mu}\bar{\Delta})\nabla_{\mu}\right].
\end{equation}
Integrating both sides of \eqref{gap} over $\tau$ and $x$ using the Riemannian volume element, we get
\begin{align}
   \frac{\beta V}{u_{0}}&= \sum_{j=-\infty}^{\infty}\textrm{Tr}\,\left[|\Delta(x)|^{2}-h_{1}(i\omega_j)\bar{\Delta}^{-1}(x)h_{2}(i\omega_j)\bar{\Delta}(x)\right]^{-1}\label{gap_hs}\\
   &=\int_{0}^{\infty}ds\, \sum_{j=-\infty}^{\infty}\textrm{Tr}\,e^{-s\left[|\Delta(x)|^{2}-h_{1}(i\omega_j)\bar{\Delta}^{-1}(x)h_{2}(i\omega_j)\bar{\Delta}(x)\right]}\label{gap_hsnext}.
\end{align}
Here, $\textrm{Tr}$ stands for the trace over the spatial variable $x$ and $V$ is the volume of the system, which is assumed to be large in the thermodynamic limit.

\subsection{Local Density Approximation in Slowly Varying Geometry}

We assume that the geometry varies slowly and consider a locally uniform gap function. The underlying idea is similar to that of the local density approximation used in the analysis of superconductivity in slowly varying trapping potentials \cite{KohnSham1965, Perali2003, Perali2004a, Perali2004b, GiorginiPitaevskiiStringari2008, Bulgac2007}. More precisely, we let $\ell_g$  be the characteristic length scale of the metric and assume that the geometry does not vary appreciably over the healing length $\ell_{h}=\hbar/\sqrt{2m \mu}$ of the system. Thus, we consider the regime $V^{1/3}\gg \ell_{g} \gg \ell_{h}$, where $V$ is the volume of the system. We then divide the volume $V$ in sufficiently small parts, within each of which the system is treated as homogeneous with a constant gap. Having established the existence of an approximate locally constant solution, we determine it by solving the integrated gap equation (\ref{gap_hs}). Working with the integrated equation allows us to extract the curvature effects via the heat kernel coefficients of a suitable single-particle operator, defined below. Finally, we restore the position dependence by allowing the locally constant solution to vary smoothly across the manifold through the curvature dependence reflected in the heat kernel coefficients, thereby obtaining a slowly varying gap function.

So, under the assumption of a real, constant gap,
\begin{equation}
\Delta(x)=\Delta=\text{const.}
\end{equation}
the operator appearing in \eqref{operator} simplifies to
\begin{equation}
|\Delta|^{2}-h_{1}(i\omega_j)\bar{\Delta}^{-1}h_{2}(i\omega_j)\bar{\Delta}
\;\longrightarrow\;
\Delta^{2}-h_{1}(i\omega_j)h_{2}(i\omega_j).
\end{equation}
and the integrated form (\ref{gap_hs}) of the saddle point equation becomes an equation for $\Delta$,
\begin{equation}
 \frac{\beta V}{u_{0}}=\sum_{j=-\infty}^{\infty}\textrm{Tr}\,\left[\Delta^{2}-h_{1}(i\omega_j)h_{2}(i\omega_j)\right]^{-1}.
\end{equation}
Choosing a basis consisting of products of temporal oscillations at the fermionic Matsubara frequencies $\omega_j$ and the eigenstates $|\sigma\rangle$ of the
single-particle operator
\begin{equation}\label{Koperator}
     K =- \frac{\hbar^2}{2m} (\nabla^{2}-\xi R),
\end{equation}
\begin{equation}
K|\sigma\rangle=\epsilon_{\sigma}|\sigma\rangle ,
\end{equation}
the operators $h_{1}$ and $h_{2}$ become diagonal and satisfy
\begin{equation}
h_{1}(i\omega_j)h_{2}(i\omega_j)\;\mapsto\;
-(\hbar\omega_j)^2-(\epsilon_{\sigma}-\mu)^2 .
\end{equation}
Therefore, the spectrum of the operator in \eqref{gap_hs} is given by
\begin{equation}
\Delta^{2}+(\hbar\omega_j)^2+(\epsilon_{\sigma}-\mu)^2 .
\end{equation}
Taking the trace as a sum over Matsubara frequencies and single-particle
states, \eqref{gap_hs} reduces to the gap equation
\begin{equation}\label{gap1}
\frac{V\beta }{u_0}=\sum_{j}\sum_{\sigma}\, \frac{1}{(\hbar \omega_{j})^{2}+\Delta^{2}(T)+(\epsilon_{\sigma}-\mu(T))^{2}}.
\end{equation}
Here and in the following, we make the temperature dependence of $\Delta$ and $\mu$ explicit.

The gap equation will be analyzed using the heat kernel methods introduced in the next section. To simplify the notation in the subsequent analysis, we adopt natural units and set $ \hbar = k_B = 1$ and $2m=1$, from this point on.

\section{Heat Kernel}
Having established our basic formalism, we now make some preliminary observations about the heat kernel expansion \cite{gilkey2004, vassilevich} on a Riemannian manifold $M$ with metric $g_{\mu\nu}$.

Following the conventions of \cite{gilkey2004} we define the heat kernel coefficients of the operator $K$ via the $t\rightarrow 0^+ $ asymptotic expansion
\begin{equation}
\textrm{Tr}\,e^{-tK}\sim \sum_{n=0}^{\infty}a_{n}(K)t^{\frac{n-d}{2}}.
\end{equation}
Here, $d$ is the dimension of the underlying manifold, which we keep general for the
moment. Since we are describing a thermodynamic system we enclose it in a box: we let $N$
be a $d$ dimensional bounded region in $M$ of volume $V$, with a smooth boundary
$\partial N$ on which a definite boundary condition (\textit{e.g.} Dirichlet, Neumann or Robin) is
imposed. The thermodynamic limit is $V\to\infty$ at fixed chemical potential, and, as we
recall below, the choice of boundary condition becomes immaterial in this limit. Following
the standard heat kernel expansion \cite{gilkey2004,vassilevich}, the first few coefficients
are given by:
\begin{align}\label{a123}
a_0(K) &= \frac{1}{(4\pi)^{d/2}} \int_N dV, \nonumber \\
a_1(K) &= \frac{\sqrt{\pi}}{2} \frac{1}{(4\pi)^{d/2}} \int_{\partial N} p \, dS, \nonumber \\
a_2(K) &= \frac{1}{(4\pi)^{d/2}} \left[ \int_N \left( \frac{1}{6} - \xi \right) R \, dV + \frac{1}{3} \int_{\partial N} L \, dS \right],
\end{align}
where $dV$ is the Riemannian volume element on $N$, $dS$ the induced surface element on
$\partial N$, and $L$ the trace of the second fundamental form (extrinsic curvature) of the
boundary. The parameter $p$ distinguishes the chosen boundary conditions, taking the value
$p=-1$ for Dirichlet and $p=+1$ for Neumann boundary conditions. The coefficients with odd
$n$, such as $a_{1}$, are pure surface terms; those with even $n$ carry a bulk
contribution --- the extensive Weyl volume term in $a_{0}$, the curvature integral in
$a_{2}$ --- accompanied, from $n=2$ on, by surface terms as well. Since bulk terms scale as
$V$ while surface terms scale as $V^{(d-1)/d}$, the latter are suppressed by a
surface-to-volume factor $V^{-1/d}$ and drop out in the thermodynamic limit, leaving only
the bulk parts of the even coefficients. In this section we nevertheless keep the general
case.
Using the heat kernel expansion  of $\textrm{Tr}\,e^{-tK}$ and the Taylor expansion of $e^{t\mu}$ we get
\begin{align}
  \textrm{Tr}\,e^{-t(K-\mu)}&= e^{t\mu}\,\textrm{Tr}\,e^{-tK}\sim \sum_{l=0}^{\infty}\sum_{k=0}^{\infty}\frac{\mu^{l}}{l!}a_{k}(K)t^{\frac{k+2l-d}{2}}\nonumber\\
  &\sim  \sum_{n=0}^{\infty}\sum_{2l\leq n} \frac{\mu^{l}}{l!}a_{n-2l}(K)\,t^{\frac{n-d}{2}}.
\end{align}
Thus,
\begin{equation}
  a_{n}(K-\mu)=\sum_{l=0}^{\left[\frac{n}{2}\right]} \frac{\mu^{l}}{l!}a_{n-2l}(K).
\end{equation}
On the other hand, from the spectral relations of higher-order elliptic operators \cite{gilkey2004}, we have
\begin{equation}\label{ankm}
  a_{n}((K-\mu)^{2})=\frac{1}{2}\frac{\Gamma\left(\frac{d-n}{4}\right)}{\Gamma\left(\frac{d-n}{2}\right)}a_{n}(K-\mu)
\end{equation}
Hence, we find
\begin{equation}\label{heat}
  \textrm{Tr}\,e^{-t(K-\mu)^{2}} \sim  \frac{1}{2}\sum_{n=0}^{\infty}\sum_{l=0}^{[n/2]}\frac{\Gamma\left(\frac{d-n}{4}\right)}{\Gamma\left(\frac{d-n}{2}\right)}\frac{\mu^{l}}{l!}
  a_{n-2l}(K)t^{(n-d)/4}.
\end{equation}
Here $[x]$ denotes the integer part of $x$.  Applying the double summation formula
\begin{equation}
  \sum_{n=0}^{\infty}\sum_{l=0}^{[n/2]}c_{l,n}=\sum_{l=0}^{\infty}\sum_{n=2l}^{\infty}c_{l,n}=\sum_{l=0}^{\infty}\sum_{n=0}^{\infty}c_{l,n+2l},
\end{equation}
we obtain
\begin{equation}\label{heatker}
  \textrm{Tr}\,e^{-t(K-\mu)^{2}} \sim  \sum_{n=0}^{\infty}\sum_{l=0}^{\infty}\frac{1}{2}\frac{\Gamma\left(\frac{d-n-2l}{4}\right)}{\Gamma\left(\frac{d-n-2l}{2}\right)}\frac{(\mu\sqrt{t})^{l}}{l!}
  a_{n}(K)t^{(n-d)/4}.
\end{equation}
For convenience let us introduce the notation
\begin{equation}\label{fn}
f_{n,d}(\mu\sqrt{t})=\frac{1}{2}\sum_{l=0}^{\infty}\frac{\Gamma\left(\frac{d-n-2l}{4}\right)}{\Gamma\left(\frac{d-n-2l}{2}\right)}\frac{(\mu\sqrt{t})^{l}}{l!}.
\end{equation}
 then
\begin{equation}\label{heatker1}
\textrm{Tr}\,e^{-t(K-\mu)^{2}}  \sim \sum_{n=0}^{\infty} f_{n,d}(\mu\sqrt{t})a_{n}(K)t^{(n-d)/4}.
\end{equation}

So we see that the heat kernel expansion is organized as a double series in powers of $\mu\sqrt{t}$ and $t$. If we combine the powers of $\sqrt{t}$ and $t$, we also see that the explicit leading terms in the heat kernel expansion are given as
\begin{align}\label{leading}
  \textrm{Tr}\,e^{-t(K-\mu)^{2}} &\sim \frac{1}{2}\frac{\Gamma\left(\frac{d}{4}\right)}{\Gamma\left(\frac{d}{2}\right)}a_{0}(K)t^{-d/4}+
   \frac{1}{2}\frac{\Gamma\left(\frac{d-1}{4}\right)}{\Gamma\left(\frac{d-1}{2}\right)}a_{1}(K)t^{-(d-1)/4}\nonumber\\
  &\quad+\frac{1}{2}\frac{\Gamma\left(\frac{d-2}{4}\right)}{\Gamma\left(\frac{d-2}{2}\right)}a_{2}(K)t^{-(d-2)/4}+\mathcal{O}(t^{(3-d)/4}).
\end{align}

As shown in \appref{AppendixA} the series \eqref{fn} defining $f_{n,d}(\mu\sqrt{t})$ can in fact be summed to give
\begin{equation}\label{sum}
  f_{n,d}(\mu\sqrt{t}) =2^{\frac{n-d}{4}}\, e^{-\frac{\mu^{2}t}{2}}D_{\frac{n-d}{2}}(-\sqrt{2}\mu \sqrt{t})
\end{equation}
Here $D_{p}(z)$ is the parabolic cylinder function whose explicit form is given in \appref{AppendixA} \cite{nist, gradshteyn}.

From the integral representation of the parabolic cylinder function $D_{p}(z)$ valid for $\textrm{Re}\,p<0$ 
\begin{equation}
  D_{p}(z)=\frac{e^{-\frac{z^{2}}{4}}}{\Gamma(-p)}\int_{0}^{\infty}dy\,y^{-p-1}\,e^{-\frac{y^{2}}{2}-yz},
\end{equation}
we immediately get for $n<d$
\begin{align}\label{fnd}
 f_{n,d}(\mu\sqrt{t})  &= \frac{2^{\frac{n-d}{4}}}{\Gamma\left(\frac{d-n}{2}\right)}\int_{0}^{\infty}dy\, y^{\frac{d-n}{2}-1}e^{-\left(\frac{y}{\sqrt{2}}-\mu\sqrt{t}\right)^{2}}\nonumber \\
 &=\frac{(\mu^{2}t)^{\frac{d-n}{4}}}{\Gamma\left(\frac{d-n}{2}\right)}\int_{0}^{\infty}dx\, x^{\frac{d-n}{2}-1}e^{-t\mu^{2}(x-1)^{2}},
\end{align}
where in the last line we made the substitution $y=\sqrt{2t\mu^{2}}x$. 

In what follows, we are going to work in three dimensions, $d=3$, and, for notational ease, set $f_{n}(\mu\sqrt{t}) =f_{n,3}(\mu\sqrt{t}) $.

\section{Gap Equation}
\subsection{Renormalization of the Gap Equation}
Recalling the gap equation \eqref{gap1} derived in Section 2, and using natural units, we can rewrite it as
\begin{equation}\label{gap_renorm}
\frac{V\beta}{u_0}=\sum_{j}\sum_{\sigma}\, \frac{1}{\omega_{j}^{2}+\Delta^{2}(T)+(\epsilon_{\sigma}-\mu(T))^{2}},
\end{equation}
where Matsubara frequencies $ \omega_{j} $ are given by
\begin{equation}
\omega_{j} =\frac{(2j+1)\pi}{\beta}.
\end{equation}
Below we will suppress the $T$ dependence of $\mu$ and $\Delta$ in our notation and we will let $\mu_{0}=\mu(0)$ and $\Delta_{0}=\Delta(0)$.

Using the Schwinger representation, 
we can rewrite the gap equation \eqref{gap_renorm} as
\begin{equation}\label{gap2}
\frac{V\beta}{u_0}=\sum_{j=-\infty}^{+\infty}\sum_{\sigma}\,\int_{0}^{\infty}\, dt\, e^{-t\left[(\omega_{j})^{2}+\Delta^{2}+(\epsilon_{\sigma}-\mu)^{2}\right] }
\end{equation}
First consider the sum over $j$ together with the Poisson summation formula, which is equivalent to Jacobi's imaginary transformation for theta functions \cite{abramowitz, gradshteyn}:
\begin{equation}
\sum_{j=-\infty}^{+\infty}e^{-(j+1/2)^{2}\pi s}= \sum_{j=-\infty}^{+\infty} \frac{1}{\sqrt{s}} (-1)^{j}\, e^{-\pi j^{2}/s}.
\end{equation}
Then
\begin{equation}
\sum_{j=-\infty}^{+\infty} e^{-t\,\omega_{j}^{2}}=\sum_{j=-\infty}^{+\infty} \frac{\beta}{\sqrt{4\pi t}} (-1)^{j}\,e^{-j^{2}\beta^{2}/4t}.
\end{equation}
If we insert the above expression into \eqref{gap2} and express the $ \sigma $ sum as the trace of the heat kernel we get
\begin{align}\label{gap3}
\frac{V}{u_0}&=\frac{1}{2\sqrt{\pi}} \int_{0}^{\infty} dt\, t^{-1/2}\, \text{Tr}\, [e^{-t[(K-\mu)^{2}+\Delta^{2}]}]\nonumber\\
&\quad+\sum_{j=1}^{\infty}\frac{(-1)^{j}}{\sqrt{\pi}} \int_{0}^{\infty} dt\, t^{-1/2}\, e^{-j^{2}\beta^{2}/4t}\,\, \text{Tr}\, e^{-t[(K-\mu)^{2}+\Delta^{2}]}
\end{align}
where $ K $ is the single-particle operator (\ref{Koperator}).

The exponential decay of the heat kernel for large $t$ ensures that there is no divergence coming from the upper limit of the integral. On the other hand, from the heat kernel expansion of a positive elliptic operator of order $4$ \cite{gilkey2004, vassilevich}
\begin{equation}\label{hk}
  \textrm{Tr}\,e^{-t[(K-\mu)^{2}+\Delta^{2}]}\sim \sum_{k=0}^{\infty}a_{k}((K-\mu)^{2}+\Delta^{2})t^{(k-3)/4},\;\;\;\;t\rightarrow 0^+.
\end{equation}
we see that there are possible divergences coming from the lower limit of the integral. For $j>0$, thanks to the $e^{-j^{2}\beta^{2}/4t}$ term, the integral is convergent at the lower limit. However for $j=0$ the exponential term is missing and the integral diverges. Note that the $j=0$ term is the zero-temperature contribution to the gap equation, and the above observation is in accordance with the general fact that in finite-temperature field theory, renormalization is needed only for the zero-temperature contributions \cite{bellac, kapusta}. 

For notational ease, let us define $L=(K-\mu)^{2}+\Delta^{2}$. Then using (\ref{leading}) and neglecting the boundary contributions in the thermodynamic limit, we render $j=0$ term finite by the subtraction
\begin{eqnarray}
 \mathcal{R}(\mu,\Delta,\Lambda)&=& \int_{0}^{\infty}\frac{dt}{\sqrt{t}}\left[\textrm{Tr}\,e^{-tL}-a_{0}(L)\frac{e^{-t\Lambda^{2}}}{t^{3/4}}\right]\\
 &=&\int_{0}^{\infty}\frac{dt}{\sqrt{t}}\left[\textrm{Tr}\,e^{-t\Delta^{2}}e^{-t(K-\mu)^{2}}-a_{0}(L)\frac{e^{-t\Lambda^{2}}}{t^{3/4}}\right].
\end{eqnarray}
Here $\Lambda$ is an ultraviolet cut-off. 

Thus, the regularized gap equation takes the form
\begin{equation}\label{gapT}
  \frac{V}{u(\Lambda)}=\frac{1}{2\sqrt{\pi}}\mathcal{R}(\mu,\Delta,\Lambda)+\sum_{j=1}^{\infty}\frac{(-1)^{j}}{\sqrt{\pi}} \int_{0}^{\infty} dt\, t^{-1/2}\, e^{-j^{2}\beta^{2}/4t}\,e^{-t\Delta^{2}}\, \text{Tr} [e^{-t(K-\mu)^{2}}]
\end{equation}
Here we write the bare coupling as $u_{0}=u(\Lambda)$ to emphasize its dependence on the ultraviolet cut-off. Taking $T\rightarrow 0$ limit we get
\begin{equation}\label{gapzero}
  \frac{V}{u(\Lambda)}=\frac{1}{2\sqrt{\pi}}\mathcal{R}(\mu_{0},\Delta_{0},\Lambda).
\end{equation}

Using the heat kernel expansion \eqref{heatker1}, and noting that there is no contribution to first few heat kernel coefficients from $e^{-t\Delta^{2}}$ term,
\begin{align}
 a_{0}(L) &= a_{0}((K-\mu)^{2})=\frac{1}{2}\frac{\Gamma\left(\frac{3}{4}\right)}{\Gamma\left(\frac{3}{2}\right)} a_{0}(K)\nonumber\\
 a_{2}(L) &= a_{2}((K-\mu)^{2})=\frac{1}{2}\frac{\Gamma\left(\frac{1}{4}\right)}{\Gamma\left(\frac{1}{2}\right) }a_{2}(K), 
\end{align}
we find
\begin{align}\label{rren}
  \mathcal{R}(\mu,\Delta,\Lambda)&=\int_{0}^{\infty}\frac{dt}{\sqrt{t}}\left\{ \left[\sum_{n=even}e^{-t\Delta^{2}}f_{n}(\mu\sqrt{t})
  a_{n}(K)t^{(n-3)/4}\right] -\frac{1}{2}\frac{\Gamma\left(\frac{3}{4}\right)}{\Gamma\left(\frac{3}{2}\right)}a_{0}(K)\frac{e^{-t\Lambda^{2}}}{t^{3/4}} \right\}.
\end{align}
Upon substituting $y=\mu^{2}t$ in the integral, we obtain a series in inverse powers of $\sqrt{\mu}$:
\begin{align}\label{R}
\mathcal{R}(\mu,\Delta,\Lambda)&=\int_{0}^{\infty}\frac{dy}{\sqrt{y}}\left\{ \left[\sum_{n=even}e^{-y(\Delta/\mu)^{2}}f_{n}(\sqrt{y})
  a_{n}(K)\mu^{-(n-1)/2}y^{(n-3)/4}\right]\right.\nonumber\\&\quad\left.-\frac{1}{2}\frac{\Gamma\left(\frac{3}{4}\right)}
  {\Gamma\left(\frac{3}{2}\right)}a_{0}(K)\sqrt{\mu}\,\frac{e^{-y(\Lambda/\mu)^{2}}}{y^{3/4}} \right\}.
\end{align}
Employing the integral representation \eqref{fnd} of $f_{n}$ in terms with $n<3$ in the above sum and using the integral representation
\begin{equation}\label{intrep}
  \frac{1}{y^{3/4}} =\frac{2}{\Gamma\left(\frac{3}{4}\right)} \int_{0}^{\infty}dx\,x^{1/2}e^{-x^{2}y},
\end{equation}
in the subtracted term of \eqref{R} we express the regularized gap equation \eqref{gapzero} as
\begin{align}\label{ulam}
  \frac{V}{u(\Lambda)}
  &= \frac{a_{0}(K)}{2\Gamma\left( \frac{3}{2}\right)} \,\mathcal{I}_{0}\left( \frac{\Delta_{0}}{\mu_{0}},\frac{\Lambda}{\mu_{0}}\right) \sqrt{\mu_{0}}+\frac{a_{2}(K)}{2\Gamma\left( \frac{1}{2}\right) }\,\mathcal{I}_{2}\left( \frac{\Delta_{0}}{\mu_{0}}\right)\frac{1}{\sqrt{\mu_{0}}}\nonumber\\
  &\quad+\frac{1}{2\sqrt{\pi}}\sum_{n=4,6,\dots}
  a_{n}(K)\mu_{0}^{-(n-1)/2}\int_{0}^{\infty}\frac{dy}{\sqrt{y}} \left[f_{n}(\sqrt{y})e^{-y(\Delta_{0}/\mu_{0})^{2}}y^{(n-3)/4}\right].
\end{align}
Here, the functions $\mathcal{I}_{0}\left( \frac{\Delta}{\mu},\frac{\Lambda}{\mu}\right)$ and $\mathcal{I}_{2}\left( \frac{\Delta}{\mu}\right)$ are defined as
\begin{align}\label{Integrals}
\mathcal{I}_{0}\left( \frac{\Delta}{\mu},\frac{\Lambda}{\mu}\right)&=\int_{0}^{\infty}dx\,\sqrt{x}\left[ \frac{1}{\sqrt{(x-1)^{2}+ \left( \frac{\Delta}{\mu}\right)^{2}}}-\frac{1}{\sqrt{x^{2}+\left( \frac{\Lambda}{\mu}\right) ^{2}}}\right]\nonumber \\
\mathcal{I}_{2}\left( \frac{\Delta}{\mu}\right)&=\int_{0}^{\infty}\frac{dx}{\sqrt{x}}\,\left[ \frac{1}{\sqrt{ (x-1)^{2}+\left( \frac{\Delta}{\mu}\right)^{2} }} \right]
\end{align}
Note that the right hand side of the gap equation (\ref{ulam}) depends on $\mu_0$ and the unknown $\Delta_0/\mu_0$
In what follows we will assume $\Delta_0/\mu_0 \ll 1$, and work up to order $\mathcal{O}(\mu_{0}^{-1/2})$. 

In \appref{AppendixB}, integrals $\mathcal{I}$ are evaluated asymptotically for $\Delta/\mu \ll 1$ \cite{abramowitz}. The results are
 \begin{align}
 \mathcal{I}_{0}\left( \frac{\Delta}{\mu},\frac{\Lambda}{\mu}\right)&=\left[ 6\log 2-4- \log \left( \frac{\Delta}{\mu}\right) ^{2}-\frac{2}{3\sqrt{\pi}}\Gamma\left(-\frac{1}{4}\right) \Gamma\left( \frac{7}{4}\right)\,\sqrt{\frac{\Lambda}{\mu}}\right]+\ldots \nonumber\\
 \mathcal{I}_{2}\left( \frac{\Delta}{\mu}\right)&=\left[6\log 2-\log \left( \frac{\Delta}{\mu} \right)^{2}\right]+\ldots
 \end{align}
Here, the ellipses represent terms that are regular in the limit $\Delta/\mu\rightarrow 0$.

Thus, at zero temperature we have
\begin{align}
  \frac{V}{u(\Lambda)} &= \frac{a_{0}\,\sqrt{\mu_{0}}}{2\Gamma\left( \frac{3}{2}\right) }\,\left[ 6\log 2-4- \log \left( \frac{\Delta_{0}}{\mu_{0}}\right) ^{2} \right]      -\frac{a_{0}}{3\sqrt{\pi}\,\Gamma\left( \frac{3}{2}\right) }\,\Gamma(7/4)\Gamma(-1/4)\sqrt{\Lambda} \nonumber\\
  &\quad+\frac{a_{2}}{2\sqrt{\mu_{0}}\,\Gamma\left( \frac{1}{2}\right) }\left[6\log 2-\log \left( \frac{\Delta_{0}}{\mu_{0}} \right)^{2}\right] +\mathcal{O}(\mu_{0}^{-3/2}).
\end{align}
So after the subtraction of the divergent term we get the renormalized coupling constant as
\begin{align}\label{uI}
\frac{V}{u_{r}} &= \frac{a_{0}\,\sqrt{\mu_{0}}}{2\Gamma\left( \frac{3}{2}\right) }\,\left[ 6\log 2-4- \log \left( \frac{\Delta_{0}}{\mu_{0}}\right) ^{2} \right] \nonumber\\
   &\quad+\frac{a_{2}}{2\sqrt{\mu_{0}}\,\Gamma\left( \frac{1}{2}\right) }\left[6\log 2-\log \left( \frac{\Delta_{0}}{\mu_{0}} \right)^{2}\right] +\mathcal{O}(\mu_{0}^{-3/2}).
\end{align}
Rearranging, we arrive at the renormalized gap equation
\begin{align}
\frac{V}{u_{r}} &= \frac{1}{2}\left[ \frac{2a_{0}\sqrt{\mu_{0}}}{\sqrt{\pi}}(6\log 2-4)+\frac{a_{2}}{\sqrt{\pi}\sqrt{\mu_{0}}}\,6\log 2+\mathcal{O}(\mu_{0}^{-3/2})\right] \nonumber\\
&\quad-\frac{1}{2}\log \left( \frac{\Delta_{0}}{\mu_{0}} \right)^{2}\left[\frac{2a_{0}\sqrt{\mu_{0}}}{\sqrt{\pi}}+\frac{a_{2}}{\sqrt{\pi}\sqrt{\mu_{0}}}+\mathcal{O}(\mu_{0}^{-3/2}) \right].
\end{align}
Thus, solving for the gap parameter, we obtain
\begin{align}\label{gapzeroagain}
\frac{\Delta_{0}}{\mu_{0}}&=\exp\left[(3\log 2-2)+\frac{a_{2}}{a_{0}}\frac{1}{\mu_{0}}+\mathcal{O}(\mu_{0}^{-2})\right]\nonumber\\
&\quad\times\exp\left[-\frac{V\sqrt{\pi}}{2a_{0}\sqrt{\mu_{0}}u_{r}} \left(1-\frac{a_{2}}{2a_{0}}\frac{1}{\mu_{0}}+\mathcal{O}(\mu_{0}^{-2})\right)\right].
\end{align}
Since $a_0$ is proportional to the volume $V$, the ratio $a_{2}/a_{0}$ is proportional to the volume average of the integrand of the bulk contribution to $a_2$ (\ref{a123}). Since we assume that the geometry varies slowly, we may approximate $a_{2}/a_{0}$ by an appropriate multiple of this integrand, evaluated at an arbitrary point within a cell where $\Delta$ is taken to be constant. This allows us to express the zero-temperature gap parameter as a spatially dependent local function, 
\begin{equation}\label{local_gap}
\begin{aligned}
\frac{\Delta_{0}(x)}{\mu_{0}} &= \exp\left[(3\log 2-2) + \left(\frac{1}{6} - \xi\right)\frac{R(x)}{\mu_{0}} + \mathcal{O}(\mu_{0}^{-2}) \right] \\
&\quad \times \exp\left[ -\frac{(4\pi)^{3/2}}{u_{r}} \frac{\sqrt{\pi}}{2\sqrt{\mu_{0}}} \left( 1 - \frac{1}{2}\left(\frac{1}{6} - \xi\right) \frac{R(x)}{\mu_{0}} + \mathcal{O}(\mu_{0}^{-2}) \right) \right].
\end{aligned}
\end{equation}
Thus, we obtain the effects of the curvature on the gap function in a slowly varying background geometry.

\subsection{Gap Equation at Finite Temperature}
Let us now return to the gap equation at finite temperature. We get the regularized gap equation by subtracting from \eqref{gap3} the divergent piece coming from its zero temperature part (as we saw, finite temperature terms are finite on their own), 
\begin{align}\label{ft}
 \frac{V}{u(\Lambda)}&= \sum_{j=-\infty}^{\infty}\frac{(-1)^{j}}{2\sqrt{\pi}} \int_{0}^{\infty} dt\, t^{-1/2}\, e^{-j^{2}\beta^{2}/4t}\,e^{-t\Delta^{2}}\, \text{Tr} \,e^{-t(K-\mu)^{2}}\nonumber\\
 &\quad-\frac{1}{2\sqrt{\pi}}\,\int_{0}^{\infty} \frac{dt}{\sqrt{t}}\, \left[a_{0}(L)\frac{e^{-t\Lambda^{2}}}{t^{3/4}}\right].
 \end{align}
 As we saw in the previous section, using the relation in \eqref{intrep}, we can write the subtracted divergence as
 \begin{equation}
\frac{a_{0}(K)}{2\Gamma(3/2)}\sqrt{\mu}\int_{0}^{\infty}\, dx\,\frac{\sqrt{x}}{\sqrt{x^{2}+\left(\frac{\Lambda}{\mu} \right)^{2} }}.
 \end{equation}

Our aim is to find the critical temperature $T_{c}$, at which $\Delta(T_{c})=0$, in terms of the observable quantity $\Delta_{0}$. Using \eqref{ulam} to express the left hand side of \eqref{ft} in terms of the gap at zero temperature, $\Delta_{0}$, and setting $\Delta=0$ and $\mu=\mu_{c}=\mu(T_{c})$ on the right hand side, we get the relevant equation. 

Using the heat kernel expansion of $\text{Tr}\, e^{-t(K-\mu)^{2}}$, we expand the first term on the right hand side of \eqref{ft} as
\begin{equation}\label{expfint}
    \frac{1}{2\sqrt{\pi}} \int_{0}^{\infty} dt\, t^{-1/2}\left[1+2\sum_{j=1}^{\infty}(-1)^{j}\,e^{-j^{2}\beta^{2}/4t}\right]\,e^{-t\Delta^{2}}\sum_{n=0,2\dots}^{\infty}
    f_{n,3}(\mu\sqrt{t})a_{n}(K)t^{(n-3)/4}.
\end{equation}
 As in the zero temperature case, we will work up to order $\mathcal{O}(\mu^{-1/2})$. Using \eqref{fnd} we see that the right hand side of \eqref{ft} takes the form
\begin{equation}
   \sum_{n=0,2}\frac{a_{n}(K)}{\Gamma\left(\frac{3-n}{2}\right)}\int_{0}^{\infty}dx\,x^{\frac{3-n}{2}-1}\mu^{(3-n)/2}
   \frac{1}{2\sqrt{\pi}}\int_{0}^{\infty} \frac{dt}{\sqrt{t}}\left[1+2\sum_{j=1}^{\infty}(-1)^{j}\,e^{-j^{2}\beta^{2}/4t}\right]\,e^{-t[\Delta^{2}+\mu^{2}(x-1)^{2}]}.
\end{equation}
Now, the $t$ integral can be performed exactly using the subordination identity
\begin{equation}\label{subord}
e^{-\lambda|u|}=\frac{\lambda}{\sqrt{\pi}}\int_{0}^{\infty} dt\, t^{-1/2}\,e^{-u^{2}/4t}\,e^{-\lambda^{2}\,t}.
\end{equation}
This results in the following expression for the right hand side of \eqref{ft}
\begin{align}\label{gapTnonzero}
  &\frac{1}{2}\sum_{n=0,2}\frac{a_{n}(K)}{\Gamma\left(\frac{3-n}{2}\right)}\int_{0}^{\infty}dx\,x^{\frac{3-n}{2}-1}\mu^{(3-n)/2}\frac{1}{\sqrt{\Delta^{2}+\mu^{2}(x-1)^{2}}} \nonumber \\
  &\quad \times \left[1+2\sum_{j=1}^{\infty}(-1)^j\,e^{-j\beta\sqrt{\Delta^{2}+\mu^{2}(x-1)^{2}}}\right]\nonumber\\
   &=  \frac{1}{2}\sum_{n=0,2}\frac{a_{n}(K)}{\Gamma\left(\frac{3-n}{2}\right)}\int_{0}^{\infty}dx\,x^{\frac{3-n}{2}-1}\mu^{\frac{3-n}{2}-1}\frac{\tanh\left(\frac{\beta\mu}{2}\sqrt{\Delta^{2}/\mu^{2}+(x-1)^{2}}\right)}
   {\sqrt{\Delta^{2}/\mu^{2}+(x-1)^{2}}}.
\end{align}
Here we used the summation formula
\begin{equation}
\tanh x = 1+2\sum_{j=1}^{\infty}\,(-1)^j\,e^{-2jx}.
\end{equation}
Written explicitly, we have
\begin{align}\label{gapTnonzero2}
  \frac{V}{u(\Lambda)} &=  \frac{a_{0}(K)}{2\Gamma(3/2)}\sqrt{\mu}\mathcal{K}_{0}(\beta\mu,\Delta/\mu,\Lambda/\mu) +\frac{a_{2}(K)}{2\Gamma(1/2)}\frac{1}{\sqrt{\mu}}\mathcal{K}_{2}(\beta\mu,\Delta/\mu)+\mathcal{O}(\mu^{-3/2}),
\end{align}
where we define
\begin{align}\label{Kintegrals1}
\mathcal{K}_{0}\left(\beta\mu,\frac{\Delta}{\mu},\frac{\Lambda}{\mu}\right) &= \int_{0}^{\infty}dx\,\sqrt{x}\left[\frac{\tanh \left( \frac{\beta\mu}{2}\sqrt{\Delta^{2}/\mu^{2}+(x-1)^{2}} \right)}{\sqrt{\Delta^{2}/\mu^{2}+(x-1)^{2}}}-\frac{1}{\sqrt{x^{2}+\left(\frac{\Lambda}{\mu} \right)^{2} }}\right],\\
\label{Kintegrals2} \mathcal{K}_{2}\left(\beta\mu,\frac{\Delta}{\mu}\right)&=\int_{0}^{\infty}\frac{dx}{\sqrt{x}}\,\frac{\tanh \left(\frac{\beta\mu}{2}\sqrt{\Delta^{2}/\mu^{2}+(x-1)^{2}}\right)}{\sqrt{\Delta^{2}/\mu^{2}+(x-1)^{2}}}.
\end{align}

\subsection{Critical Temperature}

As in the usual BCS theory in flat space, $T_{c}$ is expected to be low \cite{tinkham} and $\mu_{c}$ large. In fact, the relevant regime is $\mu_{c}\gg T_{c}$. So, setting $\beta=\beta_{c}$, $\mu=\mu_{c}$, and $\Delta=0$ in \eqref{gapTnonzero2} we arrive at
\begin{align}\label{gapTc}
  \frac{V}{u(\Lambda)} &=  \frac{a_{0}(K)}{2\Gamma(3/2)}\sqrt{\mu_{c}}\mathcal{J}_{0}(\beta_{c}\mu_{c},\Lambda/\mu_{c})+\frac{a_{2}(K)}{2\Gamma(1/2)}\frac{1}{\sqrt{\mu_{c}}}\mathcal{J}_{2}(\beta_{c}\mu_{c})+\ldots,
\end{align}
where we defined
\begin{align}\label{JKrel}
  \mathcal{J}_{0}\left(\beta\mu,\frac{\Lambda}{\mu}\right)&=\mathcal{K}_{0}\left(\beta\mu,0,\frac{\Lambda}{\mu}\right),\\
  \label{JKrel2}\mathcal{J}_{2}\left(\beta\mu\right)&=\mathcal{K}_{2}\left(\beta\mu,0\right).
\end{align}

The asymptotic behavior of the $\mathcal{J}$ integrals for large values of $\beta\mu$ are discussed in \appref{AppendixC}. Here we just quote the results with $\mu=\mu_{c}$ and $\beta=\beta_{c}$
\begin{align}
\mathcal{J}_{0} &=-4+4\log 2-\frac{\Gamma(-1/4)\Gamma(7/4)}{\Gamma(1/2)}\frac{2}{3}\sqrt{\frac{\Lambda}{\mu_{c}}}+2\left\lbrace\gamma-\log\frac{\pi}{2\mu_{c}\beta_{c}}  \right\rbrace   \nonumber\\
 \mathcal{J}_{2}&= 4\log 2+2 \left\{\gamma-\log[\pi/2\beta_{c}\mu_{c}] \right\}
\end{align}

\begin{align}
  \frac{V}{u(\Lambda)} &=  \frac{a_{0}(K)}{2\Gamma(3/2)}\sqrt{\mu_{c}}\left( -4+4\log 2-\frac{\Gamma(-1/4)\Gamma(7/4)}{\Gamma(1/2)}\frac{2}{3}\sqrt{\frac{\Lambda}{\mu_c}}+2\left\lbrace\gamma-\log\frac{\pi}{2\mu_{c}\beta_{c}}  \right\rbrace\right)\nonumber\\
  &\quad+\frac{a_{2}(K)}{2\Gamma(1/2)}\frac{1}{\sqrt{\mu_{c}}}\left( 4\log 2+2 \left\{\gamma-\log[\pi/2\beta_{c}\mu_{c}] \right\}\right) \nonumber\\
  &\quad+\ldots,
\end{align}
Subtracting the divergent terms;
\begin{align}\label{uJ}
  \frac{V}{u_{r}} &=  \frac{a_{0}(K)}{2\Gamma(3/2)}\sqrt{\mu_{c}}\left( -4+4\log 2+2\left\lbrace\gamma-\log\frac{\pi}{2\mu_{c}\beta_{c}}  \right\rbrace\right)\nonumber\\
  &\quad+\frac{a_{2}(K)}{2\Gamma(1/2)}\frac{1}{\sqrt{\mu_{c}}}\left( 4\log 2+2 \left\{\gamma-\log[\pi/2\beta_{c}\mu_{c}] \right\}\right) \nonumber\\
  &\quad+\ldots,
\end{align}
This can be solved for $T_c$ with the result
\begin{align}
  T_{c} = \frac{8\mu_{c}}{\pi} e^{\gamma-2}\exp\left[-\frac{V}{u_{r}}\left(\frac{\sqrt{\pi}}{2a_{0}\sqrt{\mu_{c}}}-\frac{\sqrt{\pi} a_{2}}{4a_{0}^{2}\mu_{c}^{3/2}}\right)\right]\exp\left[\frac{a_{2}}{a_{0}\mu_{c}}\right].
\end{align}

In flat space, for $T\lesssim T_c$, $\mu(T)-\mu_0$ is exponentially small in the weak-coupling limit. Moreover, for a weakly interacting system $\mu_{0}\simeq \varepsilon_{F}^{(0)}$, where $\varepsilon_{F}^{(0)}$ is the Fermi energy of the non-interacting system. We do not expect a slowly varying geometry to invalidate these features of weakly-coupled BCS theory, and therefore take $\mu_c\simeq \mu_0$. Using this approximation in \eqref{uJ} and comparing the resulting expression with \eqref{uI} yields
\begin{equation}\label{Tc}
\Delta_{0}\simeq e^{-\gamma}\,\pi\,T_{c},
\end{equation}
which is the same relation as in the flat space case (see \textit{e.g.} \cite{stoof}).

\subsection{Approaching the Critical Temperature}

Now we return to the gap equation \eqref{gapTnonzero2} at finite temperature and examine it for $T\simeq T_{c}$. We do so by Taylor expanding the right hand side of  \eqref{gapTnonzero2} as a function of $T$ and $\Delta^{2}$  around $T=T_{c}$ and $\Delta^{2}=0$, 
\begin{eqnarray}\label{gapexpan}
  \frac{V}{u(\Lambda)} = \frac{1}{2}\sum_{n=0,2}&&\frac{a_{n}(K)}{\Gamma\left(\frac{3-n}{2}\right)}\mu_{c}^{(3-n)/2-1}\left\{\left.\mathcal{J}_{n}\right|_{T=T_{c}}+\left.\frac{\partial\mathcal{K}_{n}}{\partial \Delta^{2}}\right|_{\substack{T=T_{c}\\\Delta^{2}=0}}\Delta^{2}+\right.\nonumber\\
  &&+\left.\left[\left.\frac{\partial\mathcal{K}_{n}}{\partial T}\right|_{\substack{T=T_{c}\\\Delta^{2}=0}}+\left.\frac{\partial\mathcal{K}_{n}}{\partial \mu}\right|_{\substack{\mu=\mu_{c}\\\Delta^{2}=0}}\,\left.\frac{\partial \mu(T)}{\partial T}\right|_{\mu=\mu(T_{c})}\right](T-T_{c})+\ldots\right\}.\nonumber\\
\end{eqnarray}
Here, again we worked up to order $\mathcal{O}(\mu^{-1/2})$. Since the variation of $\mu(T)$ with respect to $T$ was already argued to be exponentially small in the weak coupling regime, we can replace $\mu_c$ with $\mu_0$ and neglect the temperature derivative of $\mu$. Then, in conjunction with the gap equation \eqref{gapTc} at $T=T_{c}$ ($\Delta_c=0$), we arrive at
\begin{equation}
 \frac{1}{2}\sum_{n=0,2}\frac{a_{n}(K)}{\Gamma\left(\frac{3-n}{2}\right)}\mu_{0}^{(3-n)/2-1}\left[\left.\frac{\partial\mathcal{K}_{n}}{\partial \Delta^{2}}\right|_{\substack{T=T_{c}\\\Delta^{2}=0}}\Delta^{2}+\left.\frac{\partial\mathcal{K}_{n}}{\partial T}\right|_{\substack{T=T_{c}\\\Delta^{2}=0}}(T-T_{c})\right]\simeq 0.
\end{equation}
The partial derivatives are readily calculated
\begin{equation}
  \left.\frac{\partial\mathcal{K}_{n}}{\partial \Delta^{2}}\right|_{\substack{T=T_{c}\\\Delta^{2}=0}}=\frac{1}{2\mu^{2}_{0}}\int_{0}^{\infty}dx\,x^{\frac{3-n}{2}-1}\left[\frac{\alpha_{c} \sech^{2}\alpha_{c}(x-1)}{(x-1)^{2}}-\frac{\tanh{\alpha_{c}}(x-1)}{(x-1)^{3}}\right],
\end{equation}
and
\begin{equation}
\left.\frac{\partial\mathcal{K}_{n}}{\partial T}\right|_{\substack{T=T_{c}\\\Delta^{2}=0}}=-\frac{\alpha_{c}}{T_{c}}\int_{0}^{\infty}dx\,x^{\frac{3-n}{2}-1}\sech^{2}\alpha_{c}(x-1).
\end{equation}
Here $\alpha_{c}=\mu_{c}/2T_{c}\simeq \mu_{0}/2T_{c}$.

As shown in \appref{AppendixD}, we find
\begin{equation}
\left.\frac{\partial\mathcal{K}_{0}}{\partial T}\right|_{\substack{T=T_{c}\\\Delta^{2}=0}}=\left.\frac{\partial\mathcal{K}_{2}}{\partial T}\right|_{\substack{T=T_{c}\\\Delta^{2}=0}}=-\frac{2}{T_{c}}
\end{equation}
and;
\begin{align}
\left.\frac{\partial\mathcal{K}_{0}}{\partial \Delta^{2}}\right|_{\substack{T=T_{c}\\\Delta^{2}=0}}&=
\left.\frac{\partial\mathcal{K}_{2}}{\partial \Delta^{2}}\right|_{\substack{T=T_{c}\\\Delta^{2}=0}}=-\frac{7}{4}\frac{\zeta(3)}{\pi^{2}\,T_{c}^{2}}.
\end{align}
Thus,
\begin{align}
 \frac{V}{u(\Lambda)}&=\frac{1}{2}\frac{a_{0}}{\Gamma(3/2)}\sqrt{\mu_{0}}\left[\mathcal{J}_{0}-\frac{7}{4}\frac{\zeta(3)}{T_{c}^{2}\pi^{2}}\Delta^{2}-\frac{2}{T_{c}}(T-T_{c}) \right] \nonumber\\
 &\quad+\frac{1}{2}\frac{a_{2}}{\Gamma(1/2)}\frac{1}{\sqrt{\mu_{0}}}\left[ \mathcal{J}_{2}-\frac{7}{4}\frac{\zeta(3)}{T_{c}^{2}\pi^{2}}\Delta^{2}-\frac{2}{T_{c}}(T-T_{c})\right].
\end{align}
Subtracting the divergent terms gives;
\begin{align}\label{uK}
 \frac{V}{u_{r}}&=\frac{1}{2}\frac{a_{0}}{\Gamma(3/2)}\sqrt{\mu_{0}}\left[-4+4\log 2+2\left\lbrace\gamma-\log\frac{\pi}{2\mu_{0}\beta_{c}}  \right\rbrace -\frac{7}{4}\frac{\zeta(3)}{T_{c}^{2}\pi^{2}}\Delta^{2}-\frac{2}{T_{c}}(T-T_{c}) \right] \nonumber\\
 &\quad+\frac{1}{2}\frac{a_{2}}{\Gamma(1/2)}\frac{1}{\sqrt{\mu_{0}}}\left[ 4\log 2+2 \left\{\gamma-\log[\pi/2\beta_{c}\mu_{0}] \right\}-\frac{7}{4}\frac{\zeta(3)}{T_{c}^{2}\pi^{2}}\Delta^{2}-\frac{2}{T_{c}}(T-T_{c})\right].
\end{align}
Comparing the gap equation at $T=T_c$ \eqref{uJ} with \eqref{uK}, we see that  the solution for $\Delta(T)$ exists only for $T\lesssim T_c$, and is given explicitly as 
\begin{equation}
\Delta(T)=\pi\,T_{c}\,\sqrt{\frac{8}{7\zeta(3)}}\,\sqrt{1-\frac{T}{T_{c}}},
\end{equation}
which, within the same approximation, can also be stated in terms of $\Delta_0$ as
\begin{equation}
\Delta(T)=e^\gamma \Delta_0\sqrt{\frac{8}{7\zeta(3)}}\,\sqrt{1-\frac{\pi T}{e^\gamma \Delta_0}}.
\end{equation}

We remark that although we obtained the relation between $\Delta(T)$ and $\Delta_0$ through expansions that relate them to the renormalized coupling constant and temperature, there is indeed a more fundamental and universal relation between them. This relation is independent of the renormalization scheme used; it can be found by taking the difference between the zero-temperature and finite-temperature results:
\begin{align}
0&=\frac{1}{2\sqrt{\pi}} \int_{0}^{\infty} dt\, t^{-1/2}\, \Big(\text{Tr}\, [e^{-t[(K-\mu(T))^{2}+\Delta(T)^{2}]}]-\text{Tr}\, [e^{-t[(K-\mu_0)^{2}+\Delta_0^{2}]}]\Big)\nonumber\\
&\quad+\sum_{n=1}^{\infty}\frac{(-1)^{n}}{\sqrt{\pi}} \int_{0}^{\infty} dt\, t^{-1/2}\, e^{-n^{2}\beta^{2}/4t}\,\, \text{Tr}\, e^{-t[(K-\mu(T))^{2}+\Delta(T)^{2}]}.
\end{align}
As we have argued above,
for a weakly interacting fermionic system, we expect $\mu(T)\approx \mu_0\approx \varepsilon_{F}^{(0)}$, in such a case we have a direct link between $\Delta(T)$ and $\Delta_0$, which is a relation among  measurable quantities, the zero temperature gap, finite temperature gap and the   temperature. One can think of $\Delta_0$  defining the microscopic  theory equally well instead  of the coupling constant $u_r$. As a consequence of this relation we show that $\frac{\partial \Delta(T)}{\partial T} <0$, within the constant chemical potential approximation.
Since the subtracted term with constant $\Delta_0$ does not depend on the temperature when we take the derivative of the whole expression with respect to $T$ the result does not contain this subtraction.
This gives us immediately,
\begin{align}
&\frac{1}{\sqrt{\pi} }\Delta(T){\partial \Delta(T)\over \partial  T} \int_{0}^{\infty} dt\, t^{1/2}\Big( 1+2\sum_{n=1}^{\infty}(-1)^{n}\,e^{-n^{2}\beta^{2}/4t}\Big) \text{Tr} [e^{-t[(K-\mu_0)^{2}+\Delta(T)^{2}]}]\nonumber\\
&=\sum_{n=1}^{\infty}\frac{(-1)^{n}}{\sqrt{\pi}} \int_{0}^{\infty} dt\, t^{-1/2}\, e^{-n^{2}\beta^{2}/4t}\frac{n^2\beta^3}{2 t}\,\, \text{Tr} e^{-t[(K-\mu_0)^{2}+\Delta(T)^{2}]}.
\end{align}

Note that the first sum on the left is again given by the expression  
\begin{align}
1+2\sum_{n=1}^{\infty}(-1)^{n}\, e^{-n^{2}\beta^{2}/4t}=\frac{2\sqrt{\pi t}}{\beta}\sum_{n=-\infty}^{\infty}e^{-t(2n+1)^{2}\pi^{2}/\beta^{2}},
\end{align}
 which is our initial formula, hence strictly positive.
 The right side can be transformed as follows, set $t=1/v$, and take the trace out of the operator relation.
 By means of subordination identity (thinking of $(K-\mu_0)^2$ in terms of the eigenvalues)  we find for the integral,
 \[
 \text{Tr} \Bigg( \sum_{n=1}^{\infty}(-1)^{n} n\beta^2e^{-n\beta\sqrt{(K-\mu_0)^2+\Delta(T)^2}} \Bigg).
 \]
 This sum can be exactly computed using the identity,
\[
 \sum_{n=1}^{\infty} (-1)^n ne^{-n\tau}=-\frac{1}{4\,\cosh^2(\tau/2)}=\frac{-\sech^2({\tau\over 2})}{4}
 \]
 giving us finally,
 \[
 \text{Tr} \Big( \sum_{n=1}^{\infty}(-1)^{n} n\beta^2e^{-n\beta\sqrt{(K-\mu_0)^2+\Delta(T)^2}} \Big)=-\frac{\beta^2}{4}\text{Tr}\Big[\sech^{2}\left(\frac{\beta}{2}\sqrt{(K-\mu_0)^2+\Delta(T)^2}\right)\Big]<0.
 \]
 Thus we show that $\frac{\partial \Delta(T)}{\partial T}<0$ as expected.

\section{Conclusions and Discussion}

We have presented a systematic analysis of the BCS gap equation for
non-relativistic fermions on a Riemannian manifold, in which the heat kernel
expansion serves simultaneously as the organizing principle of the
calculation and as the renormalization scheme. Let us summarize the main
elements of this framework and comment on their significance.

The central technical result is the reorganization of the expansion of
$\textrm{Tr}\,e^{-t(K-\mu)^{2}}$ as a double series whose coefficient
functions $f_{n,d}(\mu\sqrt{t})$ admit a closed-form summation in terms of parabolic cylinder functions, \eqref{sum}. Combined with the Poisson
resummation of the Matsubara frequencies --- which cleanly isolates the
ultraviolet divergences in the zero-temperature sector --- this converts the
renormalized gap equation into an asymptotic series in inverse powers of
$\sqrt{\mu}$ whose coefficients are the heat kernel coefficients $a_{n}(K)$
of the underlying manifold. The divergences are thereby identified once and
for all: a
power-law divergence proportional to $a_{0}(K)\sqrt{\Lambda}$ which is absorbed into the renormalized coupling $u_{r}$. All geometric information --- volume, boundary,
and curvature contributions --- then enters the physical quantities only
through the finite parts of the $a_{n}(K)$, in a manner reminiscent of the
role these coefficients play in spectral geometry and in the Weyl expansion.

The gap equation is treated by assuming that the length scale of the curvature is much larger than the healing length of the condensate, and therefore the gap function becomes locally uniform within macroscopically infinitesimal volume elements. After the calculations are performed locally, the position dependence of the gap is reintroduced. Within this scheme we computed the zero-temperature gap $\Delta_{0}$ and the
critical temperature $T_{c}$, each of which acquires geometric corrections
organized in powers of $\mu_{0}^{-1/2}$. These corrections, however, cancel
in the ratio: to the order computed, the theory reproduces the universal
relation $\Delta_{0}=\pi e^{-\gamma}\,T_{c}$ and the standard square-root
vanishing of the gap, $\Delta(T)\propto\sqrt{1-T/T_{c}}$, near the critical
temperature. We regard this as an explicit verification, on an arbitrary
Riemannian background with slowly varying geometry, of the expected
universality of the BCS mechanism: the geometry renormalizes the effective
coupling and thereby shifts $\Delta_{0}$ and $T_{c}$ individually, but leaves
the dimensionless relations between observables intact.

Perhaps the most satisfying result of our analysis is the
regularization-independent relation obtained in Section 4.4, which connects
the zero-temperature gap, the finite-temperature gap, and the temperature
without any reference to the coupling constant or to the subtraction
procedure. This relation expresses the fact that $\Delta_{0}$ may equally
well be taken as the defining parameter of the microscopic theory in place of
$u_{r}$, and it is formulated entirely in terms of traces of heat kernels ---
that is, in terms of spectral data of the single-particle operator. As a
direct consequence we obtained an analytic proof of the monotonicity
$\partial\Delta/\partial T<0$ within the constant chemical potential
approximation, valid for an arbitrary manifold, resting only on the
positivity properties of the theta-like sums involved.

Several extensions suggest themselves. First, the uniform saddle point
adopted here is the zeroth order of a gradient expansion; the fluctuations of
the condensate, which on a curved manifold are induced at order of the
inverse curvature length scale, can be incorporated perturbatively, and their
systematic treatment would yield the curved-space Ginzburg-Landau functional
with coefficients again expressible in heat kernel data. Second, the general
formulas obtained here become concrete predictions once the coefficient
$a_{2}(K)$ is evaluated for specific geometries --- such as constant-curvature spaces, where this coefficient is controlled by the Ricci scalar; the resulting shifts in $T_{c}$ are then explicitly
computable. Third, the same framework accommodates explicit
curvature-dependent terms in the Hamiltonian, which merely modify the
endomorphism entering the heat kernel coefficients without altering the
renormalization structure. We remark that, the closed-form summation of the
coefficient functions $f_{n,d}$ in terms of parabolic cylinder functions may
prove useful beyond the present context, in other problems where a chemical
potential accompanies an operator of Laplace type. Finally, as an alternative picture, the quasi-particle spectrum around our constant or curvature improved solution can be worked out by means of the Bogoliubov theory \cite{bogoliubov} and then study their contribution at finite temperatures.

\begin{appendices}
\section{}\label{AppendixA}

In this appendix we prove the formula \eqref{sum}. First note that using the duplication formula $\Gamma(x)\Gamma(x+\frac{1}{2})=2^{1-2x}\sqrt{\pi}\Gamma(2x)$ for the gamma function we get
\begin{align}\label{factorial}
  (2k)! &= (2k)\Gamma(2k)=(2k)\frac{2^{2k-\frac{1}{2}}}{\sqrt{2}\Gamma(1/2)}\Gamma(k)\Gamma(k+\frac{1}{2})= 2^{2k}k!\left(\frac{1}{2}\right)_{(k)}.
\end{align}
Here $(a)_{(k)}=a(a+1)\ldots (a+k-1)=\frac{\Gamma(a+k)}{\Gamma(a)}$ is the Pochhammer symbol. Similarly we also find
\begin{equation}\label{factorial2}
  (2k+1)! = 2^{2k}k!\left(\frac{3}{2}\right)_{(k)}.
\end{equation}

Now define $a=(d-n)/4$ and separate the series \eqref{fn} for $f_{n,d}(\mu\sqrt{t})$  into even and odd parts in $\mu\sqrt{t}$. The even part of the series is
\begin{equation}
  \frac{1}{2}\sum_{k=0}^{\infty} (\mu\sqrt{t})^{2k}\frac{1}{(2k)!}\frac{\Gamma(a-k)}{\Gamma(2a-2k)}.
\end{equation}
Note that
\begin{align}
  \Gamma(a)&=(a-1)\ldots(a-k)\Gamma(a-k),   \\
  \Gamma(2a) &=(2a-1)\ldots(2a-2k)\Gamma(2a-2k)\nonumber\\
  &= 2^{k}[(a-1)(a-2)\ldots (a-k)][(2a-1)(2a-3)\ldots (2a-2k+1)]\Gamma(2a-2k).
\end{align}
Thus
\begin{equation}
  \frac{\Gamma(a-k)}{\Gamma(2a-2k)}=\frac{\Gamma(a)}{\Gamma(2a)}2^{k}[(2a-1)(2a-3)\ldots (2a-2k+1)].
\end{equation}
But we also have
\begin{align}
  (2a-1)(2a-3)\ldots (2a-2k+1)&=2^{k}(-1)^{k}(-a+\frac{1}{2})(-a+\frac{3}{2})\ldots(-a+k-\frac{1}{2})\nonumber\\
  &= (-1)^{k}2^{k}\left(-a+\frac{1}{2}\right)_{(k)}
\end{align}
So
\begin{equation}\label{gammaratio}
  \frac{\Gamma(a-k)}{\Gamma(2a-2k)}=\frac{\Gamma(a)}{\Gamma(2a)}(-1)^{k}2^{2k}\left(-a+\frac{1}{2}\right)_{(k)}.
\end{equation}
Using this and \eqref{factorial} we obtain
\begin{align}\label{even}
  \frac{1}{2}\sum_{k=0}^{\infty}\frac{(\mu \sqrt{t})^{2k}}{(2k)!} \frac{\Gamma(a-k)}{\Gamma(2a-2k)}&=\frac{1}{2}\frac{\Gamma(a)}{\Gamma(2a)}\sum_{k=0}^{\infty}\frac{(-\mu^{2}t)^{k}}{k!}\frac{\left(-a+\frac{1}{2}\right)_{(k)}}{\left(\frac{1}{2}\right)_{k}}\nonumber\\
  &=\frac{1}{2}\frac{\Gamma(a)}{\Gamma(2a)} \,_{1}F_{1}(-a+\frac{1}{2},\frac{1}{2},-\mu^{2}t)\nonumber\\
  &=\frac{1}{2}\frac{\Gamma((d-n)/4)}{\Gamma((d-n)/2)}\,_{1}F_{1}\left(\frac{n-d+2}{4},\frac{1}{2},-\mu^{2}t\right).
\end{align}
Here $_{1}F_{1}$ is the confluent hypergeometric
\begin{equation}\label{conhyp}
  _{1}F_{1}(a,b,x)=\sum_{k=0}^{\infty}\frac{a_{(k)}}{b_{(k)}}\frac{x^{k}}{k!}.
\end{equation}
Now using the duplication formula for the gamma function and the Kummer's formula $_{1}F_{1}(a,b,z)=e^{z}\,_{1}F_{1}(b-a,b,-z)$ we have the even part as
\begin{equation}\label{evenpart}
  \frac{2^{\frac{n-d}{2}}\sqrt{\pi}}{\Gamma\left(\frac{1}{2}-\frac{n-d}{4}\right)}e^{-\mu^{2}t}\,_{1}F_{1}\left(-\frac{n-d}{4},\frac{1}{2},\mu^{2}t\right).
\end{equation}

On the other hand the odd part of \eqref{fn} is given by
\begin{equation}
 \frac{1}{2} \sum_{k=0}^{\infty} (\mu\sqrt{t})^{2k+1}\frac{1}{(2k+1)!}\frac{\Gamma(a-k-\frac{1}{2})}{\Gamma(2a-2k-1)}.
\end{equation}
Making the substitution $a\rightarrow a-\frac{1}{2}$ in \eqref{gammaratio} and using \eqref{factorial2} we see that the odd part can be written as
\begin{equation}\label{odd}
 \frac{1}{2}  \mu\sqrt{t}\frac{\Gamma(a-\frac{1}{2})}{\Gamma(2a-1)} \sum_{k=0}^{\infty} (\mu\sqrt{t})^{2k}\frac{1}{k!}\frac{(-a+1)_{(k)}}{(3/2)_{(k)}}=\frac{1}{2}  \mu\sqrt{t}\frac{2\Gamma(\frac{d-n}{4}-\frac{1}{2})}{\Gamma(\frac{d-n}{2}-1)}\,_{1}F_{1}\left(\frac{n-d}{4}+1,\frac{3}{2},-\mu^{2} t\right).
\end{equation}
Again using the duplication formula for the gamma function and the Kummer's formula for the confluent hypergeometric function we obtain the following expression for the odd part
\begin{equation}\label{oddpart}
  \frac{2^{\frac{n-d}{2}}2\sqrt{\pi}\mu\sqrt{t}}{\Gamma\left(-\frac{n-d}{4}\right)}e^{-\mu^{2}t}\,_{1}F_{1}\left(\frac{1}{2}-\frac{n-d}{4},\frac{3}{2},\mu^{2}t\right).
\end{equation}

So adding \eqref{evenpart} and \eqref{oddpart} we arrive at.
\begin{equation}
 f_{n,d}(\mu\sqrt{t}) =2^{\frac{n-d}{4}}\,e^{-\frac{\mu^{2}t}{2}}D_{\frac{n-d}{2}}(-\sqrt{2}\mu\sqrt{t}).
\end{equation}
Here $D_{p}(z)$ is the parabolic cylinder function\cite[Eq.~9.240]{gradshteyn}:
\begin{equation}
  D_{p}(z)=2^{\frac{p}{2}}e^{-\frac{z^{2}}{4}}\left[\frac{\sqrt{\pi}}{\Gamma\left(\frac{1}{2}-\frac{p}{2}\right)}\, _{1}F_{1}\left(-\frac{p}{2},\frac{1}{2},\frac{z^{2}}{2}\right)
  -\frac{\sqrt{2\pi}z}{\Gamma\left(-\frac{p}{2}\right)}\,_{1}F_{1}\left(\frac{1}{2}-\frac{p}{2},\frac{3}{2},\frac{z^{2}}{2}\right)\right].
\end{equation}

\section{}\label{AppendixB}
In this appendix we show the asymptotic evaluation of integrals in \eqref{Integrals}  for small $\Delta/\mu$. We start with the following integral which is elementary,
\begin{align}
\mathcal{I}_{1}\left( \frac{\Delta}{\mu},\frac{\Lambda}{\mu}\right) &=\lim_{L\rightarrow \infty}\int_{0}^{L}dx\,\left[ \frac{1}{\sqrt{(x-1)^{2}+\left( \frac{\Delta}{\mu}\right) ^{2} }}-\frac{1}{\sqrt{x^{2}+\left( \frac{\Lambda}{\mu}\right) ^{2}}}\right]\nonumber\\
&=\lim_{L\rightarrow\infty}\log \left[ \frac{L-1+\sqrt{(L-1)^2+\left( \frac{\Delta}{\mu}\right) ^{2}}}{\sqrt{1+\left( \frac{\Delta}{\mu}\right)^{2} }-1}\frac{\frac{\Lambda}{\mu}}{L+\sqrt{\left(\frac{\Lambda}{\mu} \right) ^{2}+L^{2}}}\right]
\end{align}
Taking the limit $L\rightarrow\infty$ we obtain
\begin{equation}
  \mathcal{I}_{1}\left( \frac{\Delta}{\mu},\frac{\Lambda}{\mu}\right)=-\log\left(\frac{-1 + \sqrt{1 + (\Delta/\mu)^2}}{(\Lambda/\mu)}\right).
\end{equation}
Assuming  $\frac{\Delta}{\mu} $ is small we find
\begin{equation}\label{as1}
\mathcal{I}_{1}\left( \frac{\Delta}{\mu},\frac{\Lambda}{\mu}\right) =\log 2+\log \frac{\Lambda}{\mu}-\log \left( \frac{\Delta}{\mu}\right) ^{2}+\ldots
\end{equation}
Here and below ellipsis stand for regular terms in $\Delta/\mu$.

Next consider
\begin{equation}
\mathcal{I}_{0}\left( \frac{\Delta}{\mu},\frac{\Lambda}{\mu}\right) =\int_{0}^{\infty}dx\,\sqrt{x}\left[ \frac{1}{\sqrt{(x-1)^{2}+ \left( \frac{\Delta}{\mu}\right)^{2}}}-\frac{1}{\sqrt{x^{2}+\left( \frac{\Lambda}{\mu}\right) ^{2}}}\right]
\end{equation}
In order to calculate this integral we add and subtract $ 1 $ from $ \sqrt{x} $ term, i.e.
\begin{align}
\mathcal{I}_{0}\left( \frac{\Delta}{\mu},\frac{\Lambda}{\mu}\right) &=\int_{0}^{L}dx\,\left[ \frac{1}{\sqrt{(x-1)^{2}+ \left( \frac{\Delta}{\mu}\right)^{2}}}-\frac{1}{\sqrt{x^{2}+\left( \frac{\Lambda}{\mu}\right) ^{2}}}\right]+\nonumber \\
&\quad+\int_{0}^{L}dx\, \frac{\sqrt{x}-1}{\sqrt{(x-1)^{2}+ \left( \frac{\Delta}{\mu}\right)^{2}}}\nonumber\\
&\quad-\int_{0}^{L}dx\,\frac{\sqrt{x}-1}{\sqrt{x^{2}+\left( \frac{\Lambda}{\mu}\right) ^{2}}}
\end{align}
Note that the first integral is just $\mathcal{I}_{1}$. The second integral is regular at $\Delta=0$ and to get the leading contribution for small $\Delta/\mu$ we can set $\Delta=0$ in it. The second integral then gives the contribution
\begin{equation}\label{as2}
 -4+2\sqrt{L}+4\log 2-2\log (1+\sqrt{L})+\ldots.
\end{equation}
Now using the identity \cite[Eq.~3.194.1]{gradshteyn}
\begin{equation}\label{hyper}
\int_{0}^{u}\,\frac{dx\,x^{a-1}}{(1+bx)^{\nu}}=\frac{u^{a}}{a}\,_{2}F_{1}(\nu,a;1+a,-bu),
\end{equation}
the third integral (with the minus sign in front) is calculated exactly as
\begin{equation}\label{hyperwL}
  -\frac{2L^{3/2}}{3(\Lambda/\mu)}\,\, _{2}F_{1}\left(\frac{1}{2},\frac{3}{4};\frac{7}{4},-\frac{L^{2}}{(\frac{\Lambda}{\mu})^{2}}\right)+\log\left(\frac{L+\sqrt{L^{2}+(\Lambda/\mu)^{2}}}{\Lambda/\mu}\right).
\end{equation}
To find the $L\to\infty$ behavior of \eqref{hyperwL}, we use the asymptotic expansion of
the hypergeometric function \cite[Eq.~15.3.7]{abramowitz}, with the hypergeometric series
on the right hand side written out to second order in $1/z$:
\begin{align}\label{gausshyper}
&_{2}F_{1}(a,b;c;z)\sim \nonumber\\
&\frac{\Gamma(b-a)\Gamma(c)}{\Gamma(b)\Gamma(c-a)}(-z)^{-a}\left(1+\frac{a(1+a-c)}{(1+a-b)z}+\frac{a(1+a)(1+a-c)(2+a-c)}{2(1+a-b)(2+a-b)z^{2}} \right)\nonumber\\
 &\quad+\frac{\Gamma(a-b)\Gamma(c)}{\Gamma(a)\Gamma(c-b)}(-z)^{-b}\left(1+\frac{b(1+b-c)}{(1-a+b)z}+\frac{b(1+b)(1+b-c)(2+b-c)}{2(1-a+b)(2-a+b)z^{2}} \right)
\end{align}
we see that the contribution of the third integral is given as
\begin{equation}\label{as3}
  -2\sqrt{L}-\frac{2}{3}\frac{\Gamma(-1/4)\Gamma(7/4)}{\sqrt{\pi}}\sqrt{\frac{\Lambda}{\mu}}+\log 2+\log L-\log\left(\frac{\Lambda}{\mu}\right).
\end{equation}
Combining \eqref{as1}, \eqref{as2} and \eqref{as3} we get the small $ \frac{\Delta}{\mu}$ asymptotic of $\mathcal{I}_{0}$ as
\begin{equation}
\mathcal{I}_{0}\left( \frac{\Delta}{\mu},\frac{\Lambda}{\mu}\right) =6\log 2-4-\log \left( \frac{\Delta}{\mu}\right) ^{2}-\frac{2}{3}\frac{\Gamma(-1/4)\Gamma(7/4)}{\sqrt{\pi}}\sqrt{\frac{\Lambda}{\mu}}.
\end{equation}

In
\begin{equation}
  \mathcal{I}_{2}=\int_{0}^{\infty}\frac{dx}{\sqrt{x}}\,\frac{1}{\sqrt{(x-1)^{2}+(\Delta/\mu)^{2}}}
\end{equation}
we substitute $ y=x-1 $
\begin{align}
\mathcal{I}_{2}\left( \frac{\Delta}{\mu}\right) &=\int_{0}^{\infty}\frac{dx}{\sqrt{x}}\,\left[ \frac{1}{\sqrt{ (x-1)^{2}+\left( \frac{\Delta}{\mu}\right)^{2} }} \right] \nonumber \\
&=\int_{-1}^{\infty}\, \frac{dy}{\sqrt{y+1}}\frac{1}{\sqrt{y^2+\left( \frac{\Delta}{\mu}\right) ^2}}
\end{align}
and divide the integral into three pieces as
\begin{equation}
\mathcal{I}_{2}\left( \frac{\Delta}{\mu}\right) =\int_{-1}^{0}+\int_{0}^{1}+\int_{1}^{\infty}.
\end{equation}
We write the integral from $x=-1$ to $x=0$ as
\begin{equation}
\int_{-1}^{0}\,\frac{dy}{\sqrt{y^2+\left( \frac{\Delta}{\mu}\right) ^2}}\left(\frac{1}{\sqrt{y+1}} -1\right)+\int_{-1} ^{0}\,dy\,\frac{1}{\sqrt{y^2+\left( \frac{\Delta}{\mu}\right) ^2}}
\end{equation}
The first integral is regular for $\Delta=0$ and it can be approximated for $\Delta \ll \mu$ by setting $\Delta=0$, whereas the second integral leads to a logarithmic term in $\Delta/\mu$. The results is
\begin{equation}\label{as4}
  3\log 2-\log \left( \frac{\Delta}{\mu}\right)+\ldots
\end{equation}
Similarly the integral from $x=0$ to $x=1$ is written as
\begin{equation}
\int_{0}^{1}\,\frac{dy}{\sqrt{y^2+\left( \frac{\Delta}{\mu}\right) ^2}}\left(\frac{1}{\sqrt{y+1}} -1\right)+\int_{0} ^{1}\,dy\,\frac{1}{\sqrt{y^2+\left( \frac{\Delta}{\mu}\right) ^2}}
\end{equation}
This is treated similarly to the previous integral and for $\Delta \ll \mu$ its contribution is
\begin{equation}\label{as5}
  -2\sinh^{-1}(1)+ 3\log 2-\log \left( \frac{\Delta}{\mu}\right)+\ldots
\end{equation}
Finally the last piece is
\begin{equation}
  \int_{1}^{\infty}\, \frac{dy}{\sqrt{y+1}}\frac{1}{\sqrt{y^2+\left( \frac{\Delta}{\mu}\right) ^2}}.
\end{equation}
This is regular at $\Delta=0$ and for $\Delta \ll \mu$ it is approximated by setting $\Delta=0$. This gives
\begin{equation}\label{as6}
  2\sinh^{-1}(1)+\ldots
\end{equation}
Combining \eqref{as4}, \eqref{as5} and \eqref{as6} we get
\begin{equation}
  \mathcal{I}_{2}\left( \frac{\Delta}{\mu}\right)=6\log2-\log\left(\frac{\Delta}{\mu}\right)^{2}.
\end{equation}

\section{}\label{AppendixC}
In this appendix we evaluate the large $\beta\mu$ expansions of $\mathcal{K}$ integrals (\ref{Kintegrals1}) and  (\ref{Kintegrals2}) at $\Delta=0$, which were denoted by $\mathcal{J}$ in (\ref{JKrel}) and (\ref{JKrel2}). We start with the integral
\begin{equation}\label{j1}
\mathcal{J}_{1}(\beta\mu,\Lambda/\mu)=\lim_{L\rightarrow \infty} \int_{0}^{L}dx\,\left[ \frac{\tanh \left[  \frac{\beta\,\mu}{2}(x-1)\right] }{x-1}-\frac{1}{\sqrt{x^{2}+\left(\frac{\Lambda}{\mu} \right)^{2} }}\right]
\end{equation}
Consider the first part
\begin{equation}
\int_{0}^{L}dx\,\frac{\tanh \left[  \frac{\beta\,\mu}{2}(x-1)\right] }{x-1}.
\end{equation}
We introduce $ \alpha=\frac{\beta\,\mu}{2} $ and shift the variable $ x $
\begin{equation}
\int_{-1}^{L-1}dy\,\frac{\tanh \left(  \alpha y\right) }{y}=\int_{0}^{1}dy\,\frac{\tanh \left( \alpha y\right) }{y}+\int_{0}^{L-1}dy\,\frac{\tanh \left(  \alpha y\right) }{y}.
\end{equation}
Integration by parts yields
\begin{equation}\label{byparts}
\int dy\,\frac{\tanh \left( \alpha y\right) }{y}=\log y\,\tanh(\alpha\,y)-\int\,dy\,\frac{\alpha\,\log y}{\cosh^{2}(\alpha y)}.
\end{equation}
Inserting the boundary values and changing the variable $z= \alpha y $ gives
\begin{equation}
\int_{-1}^{L-1}dy\,\frac{\tanh \left(  \alpha y\right) }{y}=\log[L-1]-\int_{0}^{\alpha}dz\,\frac{\log(z/\alpha)}{\cosh^{2}z}-\int_{0}^{\alpha(L-1)}dz\,\frac{\log(z/\alpha)}{\cosh^{2}z}.
\end{equation}
Using the identity \cite[Eq.~4.371.3]{gradshteyn} 
\begin{equation}\label{id}
\int_{0}^{\infty}dz\,\frac{\log(z/\alpha)}{\cosh^{2}z}=\log[\pi\,e^{-\gamma}/4\alpha],
\end{equation}
the leading contribution in the $ \alpha\rightarrow \infty $ limit is seen to be
\begin{equation}\label{J11}
\int_{-1}^{L-1}dy\,\frac{\tanh \left(  \alpha y\right) }{y}=\log[L-1]+2\left\lbrace\gamma-\log(\pi/4\alpha) \right\rbrace+\ldots
\end{equation}
Now consider the second integral in \eqref{j1}
\begin{equation}
\int_{0}^{L}dx\,\frac{1}{\sqrt{x^{2}+\left(\frac{\Lambda}{\mu} \right)^{2} }}=-\log\left( \frac{\Lambda}{\mu}\right) +\log[L+\sqrt{L^{2}+(\Lambda/\mu)^{2}}].
\end{equation}
Subtracting this from \eqref{J11} and taking the limit $L\rightarrow\infty$ we find the large $\beta\mu$ expansion
\begin{equation}
\mathcal{J}_{1}(\beta\mu,\Lambda/\mu) =\log\left( \frac{\Lambda}{\mu}\right)+2\left\lbrace\gamma-\log[\pi/2\beta\mu] \right\rbrace -\log 2 +(\ldots).
\end{equation}

Now consider $ \mathcal{J}_{0} $
\begin{equation}
\mathcal{J}_{0}(\beta\mu,\Lambda/\mu) = \int_{0}^{L}dx\,\sqrt{x}\left[\frac{\tanh \left[  \frac{\beta\,\mu}{2}(x-1)\right] }{x-1}-\frac{1}{\sqrt{x^{2}+\left(\frac{\Lambda}{\mu} \right)^{2} }}\right].
\end{equation}
Using the identity
\begin{equation}
\frac{\sqrt{x}}{x-1}=\frac{1}{\sqrt{x}+1}+\frac{1}{x-1},
\end{equation}
we find
\begin{equation}\label{J00}
\int_{0}^{L} dx\,\left[\frac{\tanh[\alpha(x-1)]}{\sqrt{x}+1}+\frac{\tanh[\alpha(x-1)]}{x-1}-\frac{\sqrt{x}}{\sqrt{x^{2}+\left(\frac{\Lambda}{\mu} \right)^{2} }}\right]
\end{equation}
Here the middle term is exactly the integral we evaluated in \eqref{J11}.

Let us focus on the first integral. Note the leading large $\alpha$ asymptotic $ \tanh(\alpha y)\simeq 1-2\Theta(-y) $, where $ \Theta $ is the step function. Then, subtracting and adding this term we get
\begin{align}
\int_{0}^{L} dx\,\frac{\tanh[\alpha(x-1)]}{\sqrt{x}+1}
&=\int_{-1}^{L-1} dy\,\frac{\tanh \alpha\, y-(1-2\Theta(-y))}{\sqrt{y+1}+1}+\int_{0}^{L} dx\,\frac{(1-2\Theta(1-x))}{\sqrt{x}+1}.
\end{align}
Now note
\begin{equation}
  \tanh \alpha y-(1-2\Theta(-y))=-2(\textrm{sgn}(y))\frac{e^{-2\alpha |y|}}{1+e^{-2\alpha |y|}},
\end{equation}
and
\begin{equation}
 \left | \frac{\tanh \alpha y-(1-2\Theta(-y))}{\sqrt{y+1}+1}\right|\leq 2 e^{-2 \alpha|y|}.
\end{equation}
Obviously $2 e^{-2 \alpha |y|}$ is integrable. Then by the dominated convergence theorem the limit can be taken inside the integral and we obtain
\begin{equation}
  \lim_{\alpha\rightarrow\infty}\int_{-1}^{L-1} dy\,\frac{\tanh \alpha y-(1-2\Theta(-y))}{\sqrt{y+1}+1}=0.
\end{equation}
On the other hand
\begin{align}\label{J01}
   \int_{0}^{L} dx\,\frac{1-2\Theta(1-x)}{\sqrt{x}+1}
   &=  \int_{0}^{L} dx\,\frac{1}{\sqrt{x}+1}-2\int_{0}^{1} dx\,\frac{1}{\sqrt{x}+1}\nonumber\\
   &= 2\sqrt{L}-2\log(1+\sqrt{L})-2(2-2\log 2).
\end{align}
Here we used the elementary result
\begin{equation}
  \int dx\,\frac{1}{\sqrt{x}+1}=2\sqrt{x}-2\log(\sqrt{x}+1).
\end{equation}
Now the last integral in \eqref{J00} in conjunction with \eqref{hyper} is given as
\begin{equation}
  \int_{0}^{L}dx\,\frac{\sqrt{x}}{\sqrt{x^{2}+(\Lambda/\mu)^{2}}}=\frac{2L^{3/2}}{3}\frac{\mu}{\Lambda}\, _{2}F_{1}\left(\frac{1}{2},\frac{3}{4},\frac{7}{4},-\frac{L^{2}\mu^{2}}{\Lambda^{2}}\right).
\end{equation}
For large $L$, using \eqref{gausshyper}, we obtain
\begin{equation}\label{J02}
  \int_{0}^{L}dx\,\frac{\sqrt{x}}{\sqrt{x^{2}+(\Lambda/\mu)^{2}}}=2\sqrt{L}+\frac{2}{3}\frac{\Gamma(-1/4)\Gamma(7/4)}{\Gamma(1/2)}\,\sqrt{\frac{\Lambda}{\mu}}+\ldots.
\end{equation}
Now combining \eqref{J01}, \eqref{J11} and \eqref{J02} and taking the limit $L\rightarrow\infty$ we find the large $\beta\mu$ asymptotic
\begin{equation}
\mathcal{J}_{0}(\beta\mu,\Lambda/\mu) =-4+4\log 2+2\left\lbrace\gamma-\log[\pi/2\beta\mu] \right\rbrace-\frac{2}{3}\frac{\Gamma(-1/4)\Gamma(7/4)}{\Gamma(1/2)}\,\sqrt{\frac{\Lambda}{\mu}}+\ldots.
\end{equation}

Finally we consider,
\begin{align}
\mathcal{J}_{2}(\beta\mu)&=\int_{0}^{L}\frac{dx}{\sqrt{x}}\,\frac{\tanh \left[  \frac{\beta\,\mu}{2}(x-1)\right] }{x-1}=\int_{-1}^{L-1}dy\,\frac{1}{\sqrt{y+1}}\,\frac{\tanh \left(  \alpha y\right) }{y}.
\end{align}
We divide the integral into three pieces as
\begin{equation}\label{3}
  \int_{-1}^{\infty}dy\ldots= \int_{-1}^{0}dy\ldots+ \int_{0}^{1}dy\ldots+ \int_{1}^{\infty}dy\ldots
\end{equation}
and write the integral from $y=-1$ to $y=0$ as
\begin{equation}
 \int_{-1}^{0}dy\,\left(\frac{1}{\sqrt{y+1}}-1\right)\,\frac{\tanh \left(  \alpha y\right) }{y}+ \int_{-1}^{0}dy\,\frac{\tanh \left(  \alpha y\right) }{y}.
\end{equation}
The integrand of the first integral is positive, monotone increasing as a function of $\alpha$, and pointwise convergent almost everywhere as $\alpha\rightarrow\infty$. Therefore by the monotone convergence theorem, $\alpha\rightarrow\infty$  can be taken inside the integral. Using $\tanh \left(  \alpha y\right)\sim 1-2\Theta(-y)$  for large $\alpha$ the result is
\begin{equation}
  \int_{-1}^{0}dy\,\left(\frac{1}{\sqrt{y+1}}-1\right)\,\frac{1-2\Theta(-y) }{y}=2\log 2.
\end{equation}
On the other hand using \eqref{byparts} we get
\begin{align}
 \int_{-1}^{0}dy\,\frac{\tanh \left( \alpha y\right) }{y}&=\int_{-\alpha}^{0}du\,\frac{\tanh u }{u}\nonumber\\
  &= \left.\log u\tanh u\right|_{0}^{\alpha}-\int_{0}^{\alpha}du\frac{\log u}{\cosh^{2}u}.
  \end{align}
 For large $\alpha$ we use \eqref{id} and approximate this expression as
\begin{equation}
  \log\alpha+\gamma-\log(\pi/4).
\end{equation}
Thus for large $\alpha$
\begin{equation}\label{jj01}
 \int_{-1}^{0}\frac{dy}{\sqrt{y+1}}\,\frac{\tanh \left(  \alpha y\right) }{y}=2\log 2+  \left\{\gamma-\log(\pi/ 4\alpha) \right\}+\ldots.
\end{equation}
Similarly, for the second integral in \eqref{3} we get
\begin{equation}\label{jj02}
 \int_{0}^{1}\frac{dy}{\sqrt{y+1}}\,\frac{\tanh \left(  \alpha y\right) }{y}=-2\sinh^{-1}(1)+2\log 2+  \left\{\gamma-\log(\pi/4\alpha) \right\}+\ldots.
\end{equation}
Finally the monotone convergence theorem can also be applied to the last integral in \eqref{3}
\begin{equation}
  \int_{1}^{\infty}dy\,\frac{\tanh \left(  \alpha y\right) }{y\sqrt{y+1}}
\end{equation}
with the result
\begin{equation}\label{jj03}
 \int_{1}^{\infty}dy\,\frac{1-2\Theta(-y) }{y\sqrt{y+1}}=2\sinh^{-1}(1)
\end{equation}
So putting \eqref{jj01}, \eqref{jj02} and \eqref{jj03} together we find the large $\beta\mu$ asymptotic of $\mathcal{J}_{2}$:
\begin{equation}
  \mathcal{J}_{2}(\beta\mu)=4\log 2+2 \left\{\gamma-\log[\pi/2\beta\mu] \right\}+\ldots.
\end{equation}

\section{}\label{AppendixD}

In this section, we evaluate the derivatives of $ \mathcal{K} $ integrals (\ref{Kintegrals1}) and  (\ref{Kintegrals2}), which are used in the expansion (\ref{gapexpan}) of the gap equation. We begin with
\begin{equation}
\mathcal{K}_{1}\left(\beta\mu,\frac{\Delta}{\mu},\frac{\Lambda}{\mu}\right) =\int_{0}^{\infty}dx\,\left[ \frac{\tanh   \frac{\beta\mu}{2}\sqrt{\Delta^{2}/\mu^{2}+(x-1)^{2}} }{\sqrt{\Delta^{2}/\mu^{2}+(x-1)^{2}}}-\frac{1}{\sqrt{x^{2}+\left(\frac{\Lambda}{\mu} \right)^{2} }}\right]
\end{equation}
whose derivative is given as
\begin{equation}
 \left.\frac{\partial\mathcal{K}_{1}}{\partial \Delta^{2}}\right|_{\substack{T=T_{c}\\\Delta^{2}=0}}=
\frac{1}{2\mu^{2}_{0}}\int_{0}^{\infty}dx\,\left[\frac{\alpha_{c} \sech^{2}\alpha_{c}(x-1)}{(x-1)^{2}}-\frac{\tanh{\alpha_{c}} (x-1)}{(x-1)^{3}}\right]
\end{equation}
Changing the variables as $x-1=y$ and then setting $\alpha_{c}y=u$, the domain of
integration extends to $(-\infty,\infty)$ as $\alpha_{c}\to\infty$. Writing, here and in the remainder of this appendix,
\begin{equation}
    \mathcal{G}(u)=\frac{\sech^{2}u}{u^{2}}-\frac{\tanh u}{u^{3}}=\frac{1}{u}\frac{d}{du}\left(\frac{\tanh u}{u}\right),
\end{equation}
 which is an even function of $u$, and
using $\alpha_{c}^{2}/\mu_{0}^{2}=\beta_{c}^{2}/4$, we get
\begin{equation}
\left.\frac{\partial\mathcal{K}_{1}}{\partial \Delta^{2}}\right|_{\substack{T=T_{c}\\\Delta^{2}=0}}
=\frac{\beta_{c}^{2}}{8}\int_{-\infty}^{\infty} du\,\mathcal{G}(u).
\end{equation}
Using the standard Matsubara frequency expansion
$\frac{\tanh u}{u}=\sum_{n=0}^{\infty}\frac{8}{(2n+1)^{2}\pi^{2}+4u^{2}}$, we have
\begin{align}
\int_{-\infty}^{\infty} du\,\mathcal{G}(u)
&= \int_{-\infty}^{\infty} du\,\frac{1}{u}\frac{d}{du}
   \left( \sum_{n=0}^{\infty} \frac{8}{(2n+1)^{2}\pi^{2} + 4u^{2}} \right) \nonumber \\
&= \sum_{n=0}^{\infty} \int_{-\infty}^{\infty} du\,
   \frac{-64}{\left((2n+1)^{2}\pi^{2} + 4u^{2}\right)^{2}}.
\end{align}
Evaluating the elementary integral over $u$ yields $-\frac{16}{(2n+1)^{3}\pi^{2}}$. Summing over the Matsubara modes naturally gives rise to the Riemann zeta function:
\begin{equation}
-\frac{16}{\pi^{2}} \sum_{n=0}^{\infty} \frac{1}{(2n+1)^{3}} = -\frac{16}{\pi^{2}} \left( 1 - \frac{1}{8} \right) \zeta(3) = -\frac{14\zeta(3)}{\pi^{2}}.
\end{equation}
Substituting this back into the derivative gives the desired result:
\begin{equation}
\left.\frac{\partial\mathcal{K}_{1}}{\partial \Delta^{2}}\right|_{\substack{T=T_{c}\\\Delta^{2}=0}}= \frac{\beta_{c}^{2}}{8} \left( -\frac{14\zeta(3)}{\pi^{2}} \right) = -\beta_{c}^{2}\frac{7\zeta(3)}{4\pi^{2}}.
\end{equation}
Now let us consider the $ n=0 $ term:
\begin{equation}
  \left.\frac{\partial\mathcal{K}_{0}}{\partial \Delta^{2}}\right|_{\substack{T=T_{c}\\\Delta^{2}=0}}=\frac{1}{2\mu^{2}_{0}}\int_{0}^{\infty}dx\,\sqrt{x}\left[\frac{\alpha_{c} \sech^{2}\alpha_{c}(x-1)}{(x-1)^{2}}-\frac{\tanh{\alpha_{c}}(x-1)}{(x-1)^{3}}\right].
\end{equation}
Adding and subtracting $1$ from $\sqrt{x}$, this can be decomposed as:
\begin{equation}\label{K0_decomp}
\left.\frac{\partial\mathcal{K}_{0}}{\partial \Delta^{2}}\right|_{\substack{T=T_{c}\\\Delta^{2}=0}}=\frac{1}{2\mu^{2}_{0}}\int_{0}^{\infty}dx\,(\sqrt{x}-1)\left[\frac{\alpha_{c} \sech^{2}\alpha_{c}(x-1)}{(x-1)^{2}}-\frac{\tanh{\alpha_{c}}(x-1)}{(x-1)^{3}}\right]+\left.\frac{\partial\mathcal{K}_{1}}{\partial \Delta^{2}}\right|_{\substack{T=T_{c}\\\Delta^{2}=0}}.
\end{equation}
Let $I$ denote the integral in the above formula. Changing the variable to $u = \alpha_c(x-1)$ gives:
\begin{equation}
I = \alpha_c^{2} \int_{-\alpha_c}^{\infty} du
    \left( \sqrt{1+\frac{u}{\alpha_c}} - 1 \right) \mathcal{G}(u) = I_1 + I_2.
\end{equation}
where $I_{1}$ and $I_{2}$ denote the contributions of the ranges $(0,\infty)$ and
$(-\alpha_{c},0)$, respectively.
Notice that $-\mathcal{G}$ is non-negative on $\mathbb{R}$. Indeed, expanding $\sinh 2u$,
\begin{equation}
-\mathcal{G}(u) = \frac{1}{u^{2}\cosh^{2}u}\left(\frac{\sinh 2u}{2u}-1\right)
= \frac{1}{u^{2}\cosh^{2}u}\sum_{k=1}^{\infty}\frac{(2u)^{2k}}{(2k+1)!} \geq 0 .
\end{equation}
To bound $I_{1}$ we use $\sqrt{1+x}-1\leq x/2$ for $x\geq0$; substituting
$x=u/\alpha_{c}$,
\begin{equation}
|I_1| \leq -\alpha_c^{2} \int_{0}^{\infty} du \left( \frac{u}{2\alpha_c} \right)\mathcal{G}(u)
= -\frac{\alpha_c}{2} \int_{0}^{\infty} du\, u\,\mathcal{G}(u) = \frac{\alpha_{c}}{2},
\end{equation}
where we used $\int_{0}^{\infty}du\,u\,\mathcal{G}(u)=\left[\tanh u/u\right]_{0}^{\infty}=-1$. Hence
$|I_{1}|$ grows at most $\mathcal{O}(\alpha_{c})$, and multiplied by the prefactor $1/(2\mu_{0}^{2})$ this
contribution scales as $\mathcal{O}(\alpha_{c}/\mu_{0}^{2})$. Since
$\left.\partial\mathcal{K}_{1}/\partial\Delta^{2}\right|_{\substack{T=T_{c}\\\Delta^{2}=0}}$
scales as $\mathcal{O}(\alpha_{c}^{2}/\mu_{0}^{2})$, the relative contribution of $I_{1}$
is strictly subdominant and vanishes in the limit $\alpha_{c}\to\infty$:
\begin{equation}
\frac{I_1}{2\mu_0^2} = \mathcal{O}\!\left(\beta_{c}^{2}\frac{1}{\alpha_c}\right)
\to 0 \quad \text{as} \quad \alpha_c \to \infty .
\end{equation}
For $I_2$, let $u = -v$ to obtain:
\begin{equation}
I_2 = -\alpha_c^{2} \int_{0}^{\alpha_c} dv
      \left( 1 - \sqrt{1-\frac{v}{\alpha_c}} \right) \mathcal{G}(v).
\end{equation}
Since $g(x)=1-\sqrt{1-x}-x$ is convex on $[0,1]$ and vanishes at both endpoints,
$g\leq0$, that is $1-\sqrt{1-x}\leq x$. Setting $x=v/\alpha_{c}$ and noting that both
factors in the integrand are positive,
\begin{equation}
|I_2| \leq -\alpha_c^{2} \int_{0}^{\alpha_c} dv \left( \frac{v}{\alpha_c} \right)\mathcal{G}(v)
\leq -\alpha_c \int_{0}^{\infty} dv\, v\,\mathcal{G}(v) = \alpha_{c}.
\end{equation}
Similarly, the relative contribution of $\frac{I_2}{2\mu_0^2} $ is subdominant and effectively vanishes compared to the leading terms as $\alpha_c \to \infty$.
Since the integral part in \eqref{K0_decomp} gives no contribution in this limit, we conclude:
\begin{equation}
\left.\frac{\partial\mathcal{K}_{0}}{\partial \Delta^{2}}\right|_{\substack{T=T_{c}\\\Delta^{2}=0}}=\left.\frac{\partial\mathcal{K}_{1}}{\partial \Delta^{2}}\right|_{\substack{T=T_{c}\\\Delta^{2}=0}}=-\beta_{c}^{2}\frac{7\zeta(3)}{4\pi^{2}}.
\end{equation}
Now let's consider the $n=2$ term;
\begin{equation}
\left.\frac{\partial\mathcal{K}_{2}}{\partial \Delta^{2}}\right|_{\substack{T=T_{c}\\\Delta^{2}=0}} = \frac{1}{2\mu^{2}_{0}}\int_{0}^{\infty}dx\,\frac{1}{\sqrt{x}}\left[\frac{\alpha_{c} \sech^{2}\alpha_{c}(x-1)}{(x-1)^{2}}-\frac{\tanh{\alpha_{c}}(x-1)}{(x-1)^{3}}\right]=\frac{1}{\mu^{2}_{0}} J
\end{equation}
where, writing $1/(2\sqrt{x})=d\sqrt{x}/dx$ and integrating by parts,
\begin{equation}
J = - \int_{0}^{\infty}dx\,\sqrt{x}\,\frac{d}{dx}\left[\frac{\alpha_{c}\sech^{2}\alpha_{c}(x-1)}{(x-1)^{2}}-\frac{\tanh\alpha_{c}(x-1)}{(x-1)^{3}}\right].
\end{equation}

Changing variables to $u=\alpha_{c}(x-1)$ and splitting
$\sqrt{1+u/\alpha_{c}}=\bigl[\sqrt{1+u/\alpha_{c}}-1-\tfrac{u}{2\alpha_{c}}\bigr]+\bigl[1+\tfrac{u}{2\alpha_{c}}\bigr]$:
\begin{equation}\label{Jsplit}
J = - \alpha_c^3 \int_{-\alpha_c}^{\infty} du
      \left( \sqrt{1+\frac{u}{\alpha_c}} - 1 - \frac{u}{2\alpha_c} \right)\frac{d\mathcal{G}}{du}
  - \alpha_c^3 \int_{-\alpha_c}^{\infty} du
      \left( 1 + \frac{u}{2\alpha_c} \right)\frac{d\mathcal{G}}{du}.
\end{equation}

The first piece of the second integral in \eqref{Jsplit} is a total derivative, and the
second is integrated by parts. Using $\mathcal{G}(-\alpha_{c})\sim-\alpha_{c}^{-3}$ and
$\bigl[u\,\mathcal{G}(u)\bigr]_{-\alpha_{c}}^{\infty}\sim-\alpha_{c}^{-2}$,
\begin{align}
-\alpha_{c}^{3}\int_{-\alpha_{c}}^{\infty}du\left(1+\frac{u}{2\alpha_{c}}\right)
\frac{d\mathcal{G}}{du}
&=\alpha_{c}^{3}\,\mathcal{G}(-\alpha_{c})
-\frac{\alpha_{c}^{2}}{2}\left\{\Bigl[u\,\mathcal{G}(u)\Bigr]_{-\alpha_{c}}^{\infty}
-\int_{-\alpha_{c}}^{\infty}du\,\mathcal{G}(u)\right\} \nonumber\\
&\sim \alpha_{c}^{2}\left[-\frac{1}{2\alpha_{c}^{2}}
+\frac{1}{2}\int_{-\alpha_{c}}^{\infty}du\,\mathcal{G}(u)\right].
\end{align}
For the remaining part of $J$ we define
\begin{equation}
I = \alpha_c^3 \int_{-\alpha_c}^{\infty} du
    \left( 1 + \frac{u}{2\alpha_c} - \sqrt{1+\frac{u}{\alpha_c}} \right)
    \frac{d\mathcal{G}}{du} = I_1 + I_2,
\end{equation}
with $I_{1}$ and $I_{2}$ again the contributions of $(0,\infty)$ and $(-\alpha_{c},0)$.
Note that $0 \leq 1 + \frac{x}{2} - \sqrt{1+x} \leq \frac{x^{2}}{8}$ for $x \geq 0$, and
that $\mathcal{G}'\geq0$ on $[0,\infty)$. Substituting $x = u/\alpha_c$ into the bound for
$I_1$:
\begin{equation}
|I_1| \leq \frac{\alpha_c}{8} \int_{0}^{\infty} du\, u^{2}\,\frac{d\mathcal{G}}{du}
= \frac{\alpha_c}{4}.
\end{equation}

$I_1$ scales at most as $\mathcal{O}(\alpha_c)$.
For $I_2$, we change the variable to $u = -v$:
\begin{equation}
I_2 = -\alpha_c^3 \int_{0}^{\alpha_c} dv
      \left( 1 - \frac{v}{2\alpha_c} - \sqrt{1 - \frac{v}{\alpha_c}} \right)
      \frac{d\mathcal{G}}{dv}.
\end{equation}
We have, $0 \leq 1 - \frac{x}{2} - \sqrt{1-x} \leq x^2$ for $0 \leq x \leq 1$. Using this bound:
\begin{equation}
|I_2| \leq \alpha_c \int_{0}^{\infty} dv\, v^{2}\,\frac{d\mathcal{G}}{dv} = 2\alpha_{c}.
\end{equation}
Combining all parts for $J$:
\begin{equation}
\frac{1}{\mu_0^2} J \sim \frac{\beta_c^2}{4} \left[ \mathcal{O}\left(\frac{1}{\alpha_c^2}\right)+ \mathcal{O}\left(\frac{1}{\alpha_c}\right) + \frac{1}{2} \int_{-\alpha_c}^{\infty} du\, \mathcal{G}(u) \right].
\end{equation}
Taking the limit $\alpha_c \to \infty$:
\begin{equation}
\left.\frac{\partial\mathcal{K}_{2}}{\partial \Delta^{2}}\right|_{\substack{T=T_{c}\\\Delta^{2}=0}}
\to \frac{\beta_c^2}{8} \int_{-\infty}^{\infty} du\, \mathcal{G}(u)
= -\beta_{c}^{2}\frac{7\zeta(3)}{4\pi^{2}}.
\end{equation}
Finally, let us consider the temperature derivatives using the generic expression. Changing variables to $u = \alpha_c(x-1)$ yields:
\begin{equation}
  \left.\frac{\partial\mathcal{K}_{n}}{\partial T}\right|_{\substack{T=T_{c}\\\Delta^{2}=0}}=-\frac{1}{T_{c}}\int_{-\alpha_{c}}^{\infty}du\,\left(1+\frac{u}{\alpha_{c}} \right) ^{\frac{3-n}{2}-1}\sech^{2}u.
\end{equation}
For very large $\alpha_{c}$, we can Taylor expand the algebraic factor:
\begin{equation}
  \left.\frac{\partial\mathcal{K}_{n}}{\partial T}\right|_{\substack{T=T_{c}\\\Delta^{2}=0}} \simeq -\frac{1}{T_{c}}\int_{-\alpha_{c}}^{\infty}du\,\left[1+\left( \frac{3-n}{2}-1\right) \frac{u}{\alpha_{c}} \right] \sech^{2}u.
\end{equation}
Taking the limit $\alpha_{c} \rightarrow \infty$, the integration domain extends to $(-\infty, \infty)$. Notice that the term proportional to $u \sech^2 u$ is an odd function and thus its integral vanishes identically over the symmetric interval. We are left with:
\begin{equation}
\left.\frac{\partial\mathcal{K}_{n}}{\partial T}\right|_{\substack{T=T_{c}\\\Delta^{2}=0}} = -\frac{1}{T_{c}}\int_{-\infty}^{\infty} du\, \sech^{2}u = -\frac{2}{T_{c}}.
\end{equation}
\end{appendices}

\end{document}